\documentclass[twocolumn,showpacs,preprintnumbers,amsmath,amssymb,prb]{revtex4}
\usepackage{graphicx}% Include figure files
\usepackage{dcolumn}% Align table columns on decimal point
\usepackage{bm}% bold math
\usepackage{color}
\usepackage{epstopdf}
\usepackage{soul}

\definecolor{gold}{rgb}{0.85,.66,0}

\begin{document}

\title{F\"{o}rster energy transfer of dark excitons enhanced by a magnetic field \\ in an ensemble of CdTe colloidal nanocrystals }
\author{Feng Liu,$^{1,2}$  A. V. Rodina,$^{3}$ D. R. Yakovlev,$^{1,3}$  A. A. Golovatenko,$^{3}$    A. Greilich,$^{1}$  E. D. Vakhtin,$^{4}$   A. Susha,$^{5}$ A. L. Rogach,$^{5}$ Yu. G. Kusrayev$^{3}$  and M. Bayer$^{1,3}$}

\affiliation{$^{1}$ Experimentelle Physik 2, Technische Universit\"at Dortmund, 44221 Dortmund, Germany}
\affiliation{$^{2}$ Department of Physics and Astronomy, University of Sheffield, Sheffield, S3 7RH, United Kingdom}
\affiliation{$^{3}$ Ioffe Institute, Russian Academy of Sciences, 194021 St. Petersburg, Russia}
\affiliation{$^{4}$ St. Petersburg State Polytechnical University, 195251 St. Petersburg, Russia}
\affiliation{$^{5}$ City University of Hong Kong, Hong Kong}

\date{\today}

\begin{abstract}
We present a systematic experimental study along with theoretical
modeling of the energy transfer in an ensemble of
closely-packed CdTe colloidal nanocrystals identified as the F\"orster resonant energy transfer
(FRET). We prove that at low temperature of 4.2 K, mainly the ground
dark exciton states in the initially excited small-size (donor)
nanocrystals participate in the dipole-dipole FRET leading to
additional excitation of the large-size (acceptor) nanocrystals. The
FRET becomes possible due to the weak admixture of the bright
exciton states to the dark states. The admixture takes place even in
zero magnetic field and allows the radiative recombination of the
dark excitons. An external magnetic field considerably enhances this
admixture, thus increasing the energy transfer rate by a factor of
$2-3$ in a field of 15~T, as well as the radiative rates of the dark
excitons in the donor and acceptor nanocrystals. The theoretical
modeling allows us to determine the spectral dependence of the
probability for the NC to serve as a donor for larger nanocrystals, to evaluate
the energy transfer rates as well as to predict their dependencies
on the magnetic field, to describe the spectral shift of the
photoluminescence maximum due to the energy transfer and to
reproduce the experimentally observed spectral dependencies of the
photoluminescence recombination dynamics in the magnetic field.
\end{abstract}


% the Physics and Astronomy.  Classification Scheme.
%73.21.-b Electron states and collective excitations in multilayers, quantum wells, mesoscopic, and nanoscale systems
%73.21.La Quantum dots
%%
%78.47.-p Spectroscopy of solid state dynamics
%78.47.jd Time resolved luminescence
%%
%78.55.-m Photoluminescence, properties and materials
%78.55.Et II-VI semiconductors
%%
%78.67.-n Optical properties of low-dimensional, mesoscopic, and nanoscale materials and structures
%78.67.Hc Quantum dots


\maketitle

\section{Introduction}
\label{SecI}

Colloidal semiconductor nanocrystals (NCs), and especially their
optical properties are attracting a lot of attention in very
different fields.\cite{Klimov2004, Rogach2008, Talapin2010} A strong
motivation here is related to promising applications ranging from
light absorbers in photovoltaics and light emitters in
optoelectronics to medicine and biology, where they can serve as
efficient luminescent markers.\cite{Rogach2008ACI, Medintz2005,
Kamat2008, Ruele2010} Due to the strong carrier confinement, NCs are appealing objects for basic research. They offer great variety of the structural parameters and engineering of the band gap profiles. It is possible to grow them as core-shell NCs of type-I or type-II band alignment, or synthesize NCs with different shapes, such as rods or platelets, which in turn can be combined in dot-in-rod or dot-in-plate structures.\cite{Donega2011} Different potential
applications of NCs are related to effects based on energy transfer in
NC ensembles \cite{Cicek2009,Franzl2004,Wang2009a} and in their various hybrid
structures \cite{Rogach2011,Chanyawadee2009,Chanyawadee2009a,Medintz2003}.
At the same time, nonradiative energy transfer in an ensemble of
closely spaced nanocrystals often leads to reduction of the
photoluminescence (PL) quantum yield as compared to samples in
solution.\cite{Bae2013}

Colloidal NCs are often based on II-VI semiconductors, e.g. CdSe,
CdS, CdTe, ZnS, and their optical properties are dominated by the
band edge excitons.\cite{Efros1996} The exciton fine structure is
controlled by the anisotropy of the crystal lattice and the nanocrystal
shape as well as by the strong electron--hole exchange interaction
that is enhanced due to the carrier confinement.\cite{Efros1996} The
lowest exciton state is optically--forbidden in electric-dipole
approximation for a one-photon process and is therefore referred to
as a "dark exciton". The optically-allowed "bright" exciton is shifted
to higher energy by the exchange energy, which can be as large as a
few meV. As a result, the recombination dynamics of NCs especially
at low temperatures is nontrivial being dependent on the population
of the dark and bright exciton states, the spin relaxation between
them and their mixing, e.g., in external magnetic fields.

In ensembles of closely packed NCs the phenomenology of
recombination dynamics becomes even richer due to the energy transfer
between the neighboring NCs. Due to the inhomogeneous broadening of the
optical transitions caused by variations in NC size and shape, the
NCs of smaller size emit in the high-energy flank of the spectrum
and can serve as donor NCs as their energy can be transferred to the
NCs of larger size (acceptor NCs), whose emission is shifted to the
lower energy side of the spectrum. Experimentally the energy
transfer is usually evidenced by the observation of spectral
diffusion (i.e., a red shift of the PL spectrum with time) or by the
observation of a spectral dependence of the photoluminescence
dynamics, where the emission decay is shortened for the donor NCs.
These findings have been reported for
CdSe~\cite{Kagan1996,Kagan1996a,Crooker2002,Achermann2003,Xu2011,Miuazaki2012,Poulikakos2014,Akselrod2014,Mork2014},
CdS~\cite{Kim2008}, CdTe~\cite{Franzl2004,Wuister2005,Osovsky2005,Lunz2010},
PbS~\cite{Rinnerbauer2008} and Si~\cite{Gusev2011} based nanostructures.
It is well established and commonly
accepted~\cite{Rogach2009,Rogach2011} that the most relevant
mechanism of energy transfer (ET) in an ensemble of NCs with average
diameter of $4-6$~nm is the F\"orster resonant energy transfer
(FRET) based on the nonradiative dipole-dipole
coupling.\cite{Foerster1948,Dexter1953,Lakowicz1999,Allan2007} The
F\"orster energy transfer rate $\Gamma_\text{ET}$ can be described
by the following equation \cite{Crooker2002,Achermann2003}:
\begin{eqnarray}
\Gamma_\text{ET} = \frac{2 \pi}{\hbar} \frac{(\mu_\text{d} \mu_\text{a} \kappa)^2}{R_\text{da}^6 n^4} \Theta .
\label{EQ:ET}
\end{eqnarray}
Here $\mu_\text{d}$ and $\mu_\text{a}$ are the transition dipole
moments of excitons in the donor and acceptor NCs, respectively,
$\kappa$ is an orientational factor (it accounts for the
distribution of angles between the donor and the acceptor dipole
moments; for random dipole orientation $\kappa^2=2/3$), $R_\text{da}$ is
the distance between donors and acceptors, $n$ is the refractive
index of the medium, and $\Theta$ is the overlap integral between
the donor emission and the acceptor absorption spectra which are
normalized to $\mu_\text{d}^2$ and $\mu_\text{a}^2$, respectively.
It can be seen from Eq.~\eqref{EQ:ET}, that $\Gamma_\text{ET}$
$\propto 1/R_\text{da}^6$ is extremely sensitive to the distance
between donor and acceptor NCs. The most efficient energy transfer
corresponds to the situation for which two nanocrystals of different
sizes are located in an immediate vicinity of each other and the ground
state emitting level of the donor NC interacts resonantly with a
higher lying absorbing level of the acceptor NC.\cite{Achermann2003}
The nonradiative dipole-dipole energy transfer represents an
additional effective channel for the decay and shortening of the PL
lifetime in the donor NCs, which is caused by the interaction of the
radiative dipoles in donor and acceptor NCs without emission and
reabsorption of real photons. The radiative part of this interaction
may also lead to energy transfer via emission and reabsorption of
real photons as well as to a radiative correction to the donor
radiative recombination rate. However, as it was shown
recently,\cite{Poddubny2015} the radiative corrections to the
radiative and energy transfer rates can be neglected for the typical
spatial separation between the donor and acceptor dipoles at which
the energy transfer is effective.

Thus, the important (dominating) role of the FRET for colloidal NCs
is well documented and has been studied by continuous-wave and
time-resolved photoluminescence. Most experiments have been
performed at room temperature
\cite{Crooker2002,Achermann2003,Gusev2011}, with some
low-temperature data being also
available\cite{Furis2005,Blumling2012,Poulikakos2014}. However, the
details of the transfer mechanism at low temperatures, when the PL
is governed by the emission from dark exciton state, as well as the
possibilities to affect the efficiency of the energy transfer by
external electric or magnetic fields, are not yet clarified and
still open for detailed investigations.

So far, the FRET from semiconductor CdSe/CdS nanorods to dye
molecules controlled by an electric field was demonstrated
\cite{Becker2006}. The responsible mechanism is based on the quantum
confined Stark effect that enable a shift of the nanorod emission
spectrum into resonance with the dye absorption band.

The effect of an external magnetic field on the FRET efficiency  in
colloidal NCs has not been studied systematically. There are two
experimental papers related to this issue. Furis et al.
\cite{Furis2005} studied CdSe NCs in high magnetic fields up to
45~T. Pronounced exciton transfer via FRET was found in the
spectrally-resolved recombination dynamics. The
magnetic-field-induced circular polarization degree of the PL was
insensitive to the FRET process and the authors suggested that this
may provide evidence for a spin-conserving FRET. The
magnetic field effect on the FRET efficiency was not discussed, no
conclusion in this respect can be drawn from the presented
experimental data. Blumling et al. \cite{Blumling2012} studied the
effect of temperature and magnetic field on the energy transfer in
CdSe NC aggregates by measuring the steady-state PL. The authors
claimed that both temperature and magnetic field can enhance the
energy transfer due to the population of the bright exciton state.

Theoretically, the dipole-dipole interaction mechanism for energy
transfer and the influence of such factors as the donor-acceptor
separation, the spectral overlap and the effect of the surrounding
\cite{Allan2007,Vincent2011,Poddubny2015} are well established. In
contrast to multipole and exchange
mechanisms,\cite{Dexter1953,Kruchinin2008} the direct dipole-dipole
coupling conserves the spin of the participating charge carries. As the
energy transfer rate depends on the dipole moments of the excitons
in the donor and acceptor NCs, the dipole-dipole energy transfer has been considered to take place only between the bright exciton
states. At the same time it is well established that the dark
exciton states in colloidal nanocrystals are activated due to an
admixture of bright exciton state wave
functions.\cite{Rodina2015} This also means that the dark
exciton states possess a nonzero dipole moment that is proportional
to the admixed bright state dipole moment $\mu$ and the radiative
life time of the dark exciton state is $\tau_F \propto 1/\mu^2$. Up
to now, the possibility of FRET between dark excitons has been
analyzed neither theoretically nor experimentally.

A quantitative analysis of the energy transfer rates from
experimental data is complicated by several factors. Usually, the
rates are extracted from the comparison of luminescence decay curves
for small NCs (donors) that participate and do not participate in
the transfer process.\cite{Crooker2002,Achermann2003,Gusev2011}
However, the donor decay curves are usually strongly nonexponential due to the
inhomogeneous spatial distribution of acceptors around the donors.
Theoretical modeling of the time evolution of donor and acceptor
populations in a mixed ensemble of CdSe NCs has been done in Refs.
\onlinecite{Kagan1996,Kagan1996a} and reproduced the observed
spectra well. However, this analysis has not taken into account the
exciton fine structure and the possibility of the dark exciton
states to contribute to the energy transfer process.

In this paper we report an experimental and theoretical study of the
energy transfer in an ensemble of closely packed CdTe NCs. The
energy transfer is evidenced experimentally by the time-dependent
shift of the PL maximum after pulsed excitation. The observed shift
is well described  theoretically by the energy transfer between small
and large NCs. An important observation concerns the time scale of
the process - at low temperatures the energy shift is observed
during times much longer than the life time of the bright exciton
that is shortened by the fast thermalization due to the relaxation to the
dark exciton state. Spectrally and temporally resolved
photoluminescence measurements reveal a strong spectral dispersion
of the exciton life time and give clear evidence of a second rise of
the PL intensity (after the initial decay) at the low energy side of
the spectrum that becomes enhanced in an applied magnetic field. Our
theoretical modeling of the donor and acceptor population takes into
account the exciton fine structure and the possibility of
dipole-dipole energy transfer from the dark exciton state. The
results of the simulations reproduce the observed emission decays of
the acceptor NCs and allow us to determine all parameters of the
energy transfer process and their dependence on the magnetic field. The
energy transfer rate from the dark exciton increases in the external
magnetic field due to the increase of the dipole moments in both the
donor and acceptor NCs.

The paper is organized as follows: Sec. \ref{SecII} gives a
description of the samples and experimental conditions.
The experimental data with an emphasis on evidencing the energy
transfer process are presented in Sec. \ref{SecIII}. The theoretical
considerations and the resulting modeling of the experimental data
are presented in Sec.  \ref{SecIV} and  Sec. \ref{SecV},
respectively. The main conclusions are summarized in Sec.
\ref{SecVI}.

%%%%%%%%%%%%%%%%%%%%%%%%%%%%%%%%%%%%%%%%%%%%%%%%%%%%%%%%%%%%%%%%%%%%%%%%%%%

\section{Experimentals}
\label{SecII}

Thiol-capped CdTe colloidal NCs were synthesized in water as
described in Ref.~\onlinecite{Rogach2007}. The exciton recombination
and spin relaxation dynamics in such NCs were reported in
Ref.~\onlinecite{Liu2014}. In this work, two samples with average
core diameters of 3.4 and 3.7~nm were studied. For the optical
experiments at cryogenic temperatures aqueous solutions of CdTe NCs
were drop-casted on a glass slice and dried. The resulting films
consist of areas of NCs with varying (high and low) in-plane
densities of NCs, which are characterized by high and low total PL
intensities, respectively. These
inhomogeneous films allow us to study and compare the effect of the
energy transfer on the ensemble PL spectrum and PL dynamics for
areas with different average spatial separation between NCs and thus
different values of $R_{da}$ determining the transfer rate.

The samples were inserted into a cryostat equipped with a 15~T
superconducting magnet. The magnetic field, \textbf{B}, was applied
in the Faraday geometry, it was oriented perpendicular to the glass
slice and parallel to the light wave vector. The sample was in
contact with helium gas and the bath temperature was varied from
$T=4.2$ up to 300~K.

Photoluminescence was excited and collected through multimode
optical fibers. The collected emission was dispersed with a 0.55-m
spectrometer. Time-integrated PL spectra were measured for
continuous-wave (CW) laser excitation with a photon energy of
3.33~eV (wavelength 372~nm) and detected with a
liquid-nitrogen-cooled charge-coupled-device camera. We denote them
as steady-state PL spectra.

For time-resolved measurements the samples were excited with
picosecond laser pulses (photon energy 3.06~eV, wavelength 405~nm,
pulse duration 50~ps, repetition frequency 150 to 500~kHz). The PL
signal was sent through the spectrometer and detected by an
avalanche photodiode (time response 50~ps) connected to a standard
time-correlated single-photon counting module. The instrument
response function of the setup was better than 800~ps. All measurements were performed at low
excitation densities of 0.1~mW/cm$^{2}$ to avoid any multiexcitonic
contribution to the emission spectra.

\section{Experimental Results}
\label{SecIII}

\subsection{Steady state PL spectra and spectrally integrated recombination dynamics }

\begin{figure}
\centering
\includegraphics[width=0.95\linewidth]{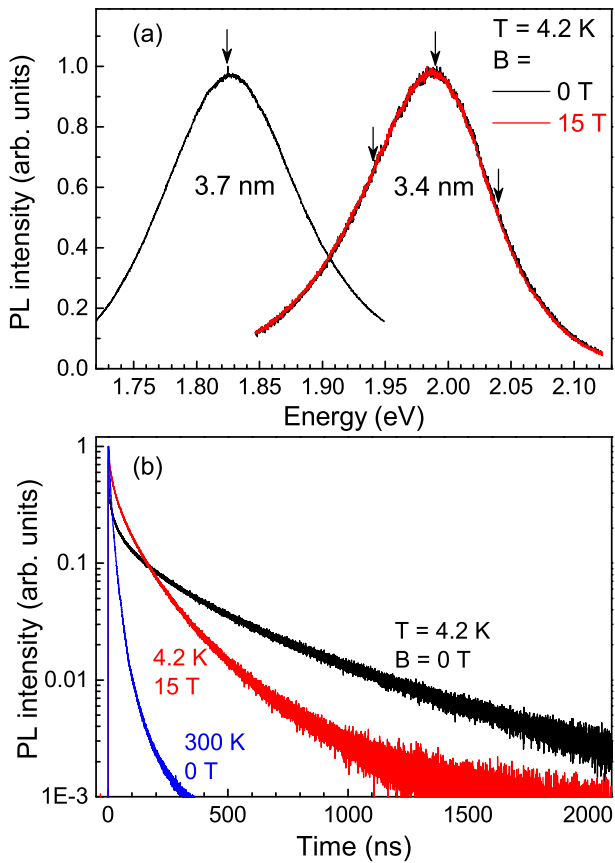}
\caption{(a) Normalized steady-state PL spectra of the 3.4~nm and
3.7~nm CdTe NCs measured at $T=4.2$~K in zero magnetic field $B=0$ T
(black lines) and at $B=15$ T (red line). The arrows indicate the
energies at which the PL dynamics shown in Figs.~\ref{fig:first50ns}
and \ref{fig:TvsB} are measured. (b) Recombination dynamics of the
spectrally-integrated PL intensity in the 3.4~nm CdTe NCs measured
at $T=4.2$~K in magnetic fields of $B=0$ T (black line) and  $B=15$
T (red line), and at $T=300$~K (blue line).} \label{fig:pl}
\end{figure}

Steady-state PL spectra of the 3.4 and 3.7~nm CdTe NCs measured
under CW excitation at $T=4.2$~K are shown in Fig.~\ref{fig:pl}(a).
Their peak positions are at 1.99~eV for the 3.4~nm NCs and 1.83~eV
for 3.7~nm the NCs. The spectra of both samples are rather broad
with a full width at half maximum (FWHM) of $\sim$120~meV, which
evidences the considerable NC size dispersion of about 7\%.
%Arrows indicate spectral positions where PL decay shown in Fig. \ref{fig:first50ns} and \ref{fig:TvsB} was measured.

The exciton recombination dynamics in the NCs can be characterized
by the spectrally integrated time-resolved PL.
Figure~\ref{fig:pl}(b) shows the integral PL decay obtained by
summing up the PL decays measured at 20 energies distributed across the
whole PL band of the 3.4~nm NCs. At room temperature the PL decay
can be described by a biexponential function with decay times of 6
and 22~ns. The longer component originates from the thermally mixed
bright and dark exciton states. With decreasing temperature down to
4.2~K the decay shows a multiexponential behavior due to the exciton
thermalization into the optically-forbidden (dark) state, which is the
lowest exciton state in NCs. It is well established that the very
fast initial decay with a time of approximately 2~ns  is related to the
optically-allowed (bright) exciton. Its decay is dominated by the
fast scattering from the bright to the dark state and has some
contribution from the radiative recombination of bright excitons
\cite{Labeau2003,Biadala2009,Blokland2011}. The slow component with
a decay time of about 260~ns corresponds to the lifetime of the dark
(optically forbidden) exciton $\tau_F$, whose recombination becomes
partially allowed due to a weak mixing of the dark and the bright
exciton states caused, e.g., by the magnetic moments of dangling
bonds at NC imperfections and surface states.

An external magnetic field induces a mixing of the bright and the
dark exciton states, which results in the vanishing of the fast decay
component and the shortening of the slow component. Such a behavior,
well established for CdSe and CdTe NCs,~\cite{Blokland2011,Liu2013}
is observed also for the studied sample. In a magnetic field of 15~T
the amplitude of the fast component ultimately vanishes and the slow
component is shortened down to about $100$~ns, compare the red and
black curves in Fig.~\ref{fig:pl}(b).

\subsection{Evidence for the energy transfer process: spectral diffusion}

\begin{figure}
\centering
\includegraphics[width=\linewidth]{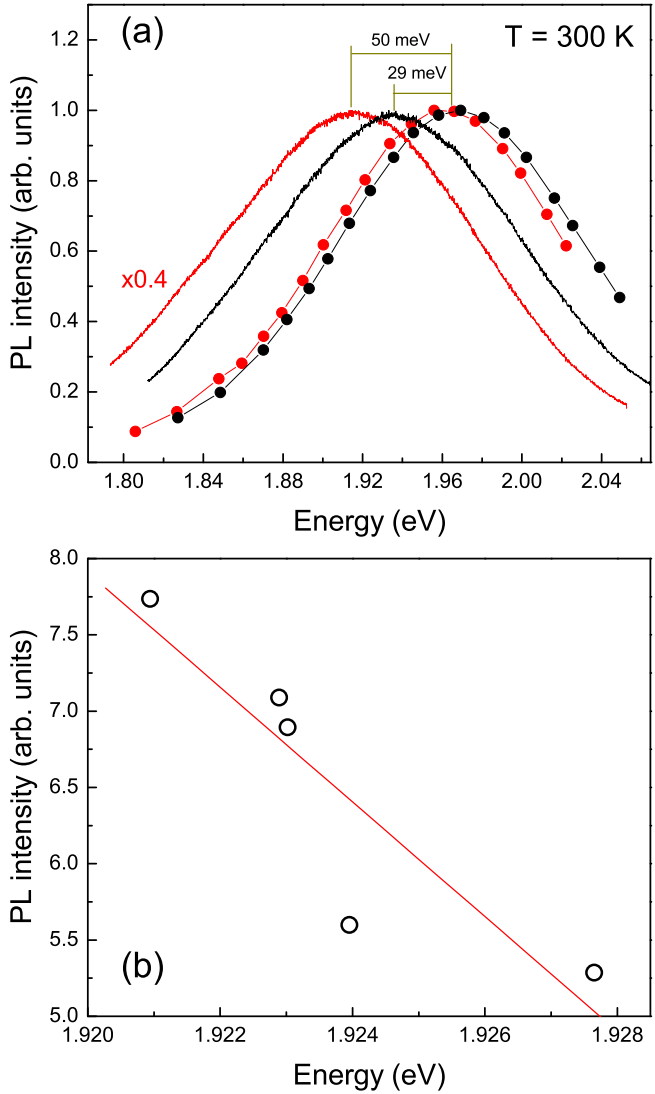}
\caption{(a) Normalized time-resolved (lines with dots) and
steady-state (solid lines) PL spectra measured at high-density (red)
and low-density (black) areas of the sample with 3.4~nm CdTe NCs.
(b) PL intensity versus peak position of the steady-state PL spectra
measured at areas with different densities of CdTe NCs. The line is
a linear interpolation.} \label{fig:compare}
\end{figure}

We now turn to the experimental results that evidence on the energy transfer in the studied CdTe NC solids.
Figure~\ref{fig:compare}(a) compares the steady-state (solid
lines) and the time-resolved (lines with dots) PL spectra of the
3.4~nm NCs measured at zero time delay just after the excitation
pulse. The red and black lines show the results measured at two
different sample areas with different NC densities, high
and low, respectively. The PL intensities at these points differ by
a factor of 2.5. The time-resolved spectra from these high- and
low-density areas are similar to each other, indicating the same NC
size dispersion in these areas. Note, that these spectra, measured
right after the pulse, are not contributed by the energy transfer
and, therefore, give us information on the density of states in the
NC ensemble.

The steady-state PL spectra in Fig.~\ref{fig:compare}(a) are shifted
to lower energies compared to the time-resolved spectra. This shift
is larger in the area with higher NC densities reaching 50~meV compared
to 29~meV in the high- and low-density areas, respectively. In
general, such shifts may be induced by the energy transfer, but also
the spectral dependence of the PL dynamics across the emission band
may be a possible origin. As the second reason is not relevant for
CdTe NCs \cite{Wuister2005}, we attribute the shift solely to the
energy transfer. The higher density of NCs corresponds to a smaller
separation between them. Therefore, for the higher NC density the
FRET is more efficient and, consequently, a lager shift between
time-resolved and steady-state PL spectra is expected.

A systematic correlation between the NC density and the energy shift
of the steady-state PL spectra measured at different sample areas is shown
in Fig.~\ref{fig:compare}(b). Here the PL intensity from the sample areas
with different NC densities is plotted against the peak energy. One
can see that stronger PL intensities correspond to the areas with
lower peak energy, i.e. larger shift. This correlation agrees with
the expectations from the F\"orster mechanism for the energy
transfer and is in line with the experimental data reported in
Ref.~\onlinecite{Lunz2010}.

\begin{figure}
\centering
\includegraphics[width=\linewidth]{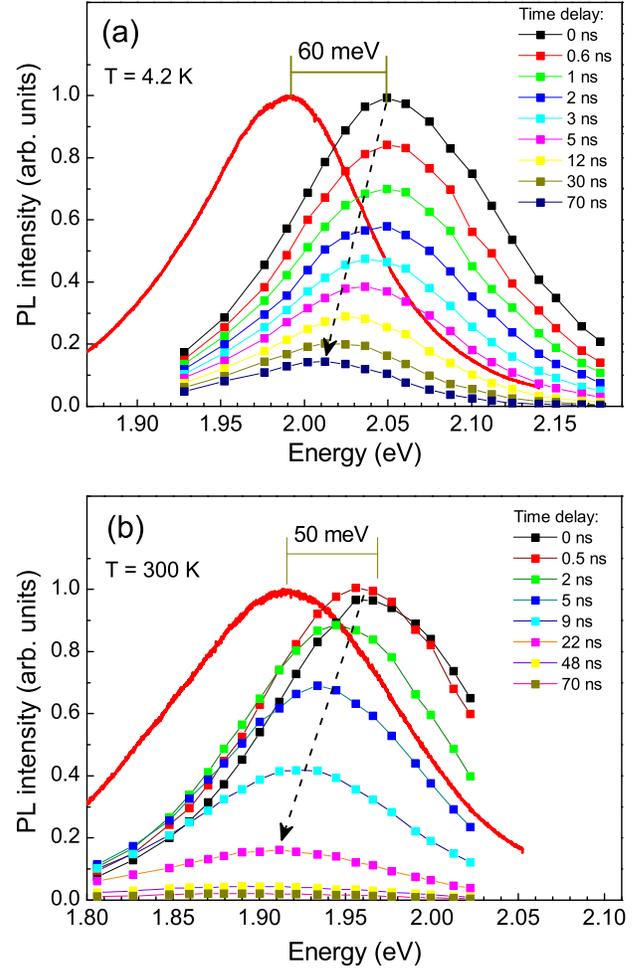}
\caption{Steady-state PL spectra (solid red lines) and time-resolved
PL spectra (lines with dots) of the 3.4~nm CdTe NCs measured at (a)
$T=4.2$~K and (b) 300~K.} \label{fig:shift}
\end{figure}

The time evolution of the PL spectra measured in high density areas
at different time delays is presented in Fig.~\ref{fig:shift} for
temperatures of 4.2 and 300~K. The strong spectral shift with
increasing delay is prominent at both temperatures. For comparison,
the steady-state PL spectra are also shown by the solid red lines.
At $T=4.2$~K the emission intensity decreases with time
monotonically and the maximum of the spectrally-resolved PL spectra
shifts from 2.05~eV at $t=0$~ns down to 2.01~eV at 70~ns, see
Fig.~\ref{fig:shift}(a). A similar behavior is observed at room
temperature, where the time-resolved spectra shift from 1.96~eV at
0~ns down to 1.92~eV at 70~ns. An important feature related to the
energy transfer is seen at room temperature in
Fig.~\ref{fig:shift}(b). Namely, the PL intensity varies
non-monotonically at the low energy tail of the emission band. It
increases during several nanoseconds after the excitation pulse and
only then starts to decay. This behavior will be shown in more
detail below where the PL dynamics measured at different spectral
energies are presented.

\begin{figure}
\centering
\includegraphics[width=\linewidth]{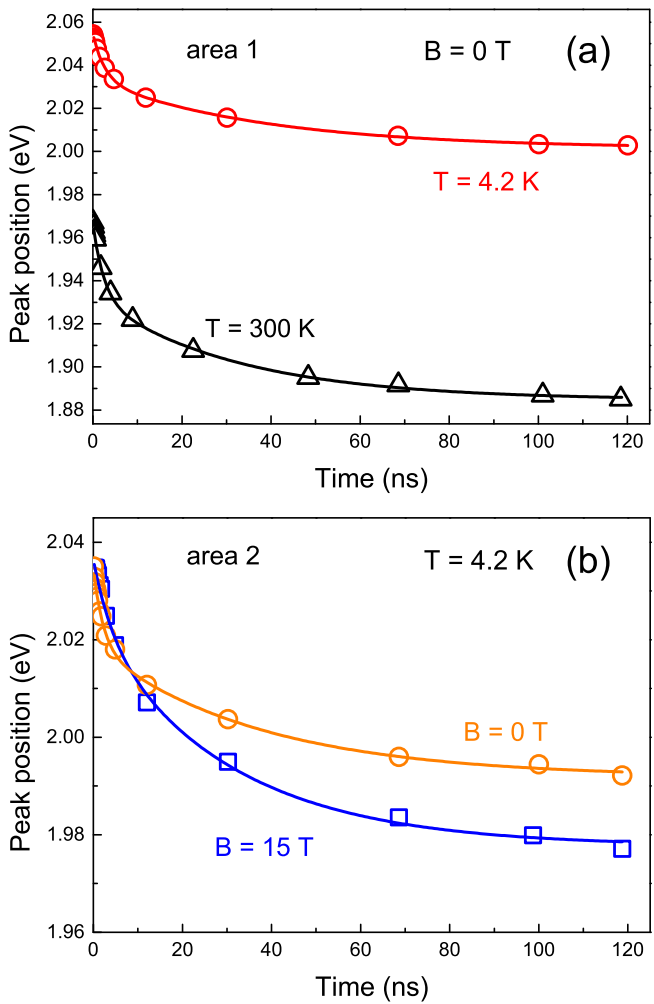}
\caption{Temporal shift of the PL maximum in 3.4~nm CdTe NCs
measured for two sample areas at different temperatures and magnetic
fields. The experimental data are shown by symbols. Lines are fits
according to Eq.~\eqref{EQ:shift1}.} \label{fig:EvsT}
\end{figure}

Figures~\ref{fig:EvsT}(a)-\ref{fig:EvsT}(c) show shifts of the PL maxima with time,
measured at two sample areas having similar PL intensities. Due to
the sample inhomogeneity several experimental data sets were
measured at different sample areas. The characteristic behavior is,
however, well reproducible for all these areas. For area 1 the PL
spectrum shifts by 47~meV during the first 70~ns at $T=4.2$~K, see
Fig.~\ref{fig:EvsT}(a). The shift becomes larger with increasing
temperature and reaches 75~meV at 300~K, indicating a more efficient
energy transfer. In both cases the fast initial energy shift during
the first 10~ns is then gradually slowed down with a tendency to
saturated. A similar shift of the PL maximum was presented in Refs.
\onlinecite{Crooker2002,Achermann2003,Akselrod2014} and attributed
to the energy transfer process at room temperature. It should be
noted, that the spectral shift (the spectral diffusion) can be also
observed in systems without any energy transfer, for example in the
donor-acceptor pair recombination in bulk
semiconductors.\cite{Thomas1965} In this case, the spectral
diffusion is caused by the spectral dependence (dispersion) of the
recombination rates. However, for CdTe NCs no dispersion of the
recombination dynamics was found in an ensemble of noninteracting
NCs, i.e. in the absence of energy transfer.\cite{Wuister2005} In
addition, we observed the same correlation between the temporal
shift of the PL maxima and the PL intensity as for the shift of the
cw spectral maxima: the areas with larger PL intensity and thus
higher NC density demonstrate a larger temporal shift of the PL
maximum. Therefore, we attribute the observed spectral shift solely
to the effect of the energy transfer.

It is important to note, that at $T=4.2$~K the bright excitons in
our NCs scatter to the dark states during 2~ns, see
Fig.~\ref{fig:pl}(b). But the energy transfer occurs at much longer
times. Therefore, it cannot be provided by the FRET involving the
bright excitons. This evidences that the dark excitons can play an
important role in the energy transfer. We will discuss this below in
more detail.

From Fig.~\ref{fig:EvsT}(b) is can be seen, that application of an
external magnetic field enhances the energy transfer. The PL
spectrum shift during the first 70~ns measured at $B=15$~T (51~meV)
is larger than that measured at 0~T (40~meV). This is in contrast to
the shift of the CW spectra which is independent of the magnetic
field, as can be seen from comparison of the CW spectra at $0$ and
$15$~T in Fig.~\ref{fig:compare}(a). Commonly, the magnetic field
enhances the carrier localization and shrinks the exciton wave
function. Therefore, the energy transfer mechanisms involving
tunneling of excitons or carriers between NCs are expected to be
slowed down in a magnetic field. But not the FRET, which can be
enhanced by, e.g., the magnetic field induced increase of the
exciton oscillator strength.

The solid lines in Fig.~\ref{fig:EvsT} are fits using a form similar
to that in Refs.~\onlinecite{Akselrod2014,Poulikakos2014}:
\begin{eqnarray}
E(t)=E_0+\Delta E_1 (e^{-t\Gamma_{\Delta E_1}}-1)+\Delta E_2 (e^{-t\Gamma_{\Delta E_2}}-1) \, .
\label{EQ:shift1}
\end{eqnarray}
The parameters of the fits, the shift energies $\Delta E_{1,2}$ and
the shift rates $\Gamma_{\Delta E_{1,2}}$ are given in Table~\ref{tab:shift}. We note that the fitting with only one
characteristic shift rate does not work well: in all cases the first
initial fast energy shift is followed by a slow shift. We comment on
the physical meaning of the different shift rates in the theory
section \ref{SecVDA}.

\begin{table}
\centering
\begin{tabular}{|c|c|c|c|c|c|}
\hline
area  & $E_0$  & $\Delta E_1$  & $\Gamma_{\Delta E_1}$  & $\Delta E_2$  & $\Gamma_{\Delta E_2}$  \\
condition &  (eV) &  (meV) &  (ns$^{-1}$) &  (meV) & (ns$^{-1}$) \\ \hline
area  1 &&&&&\\ 4.2 K, 0 T & 2.055& 22 & 0.38  & 32 & 0.025 \\ \hline
 area  1 &&&&&\\ 300 K, 0 T & 1.967& 34 & 0.41  & 48 & 0.031 \\ \hline
 area 2 &&&&&\\ 4.2 K, 0 T & 2.035 & 16 & 0.45 & 27 & 0.027  \\ \hline
area  2 &&&&&\\4.2 K, 15 T & 2.036 & 14 & 0.23 & 45 & 0.033  \\ \hline
\end{tabular}
\caption{Fit parameters for modeling the time evolution of the PL
peak maximum according to Eq.~\eqref{EQ:shift1}.} \label{tab:shift}
\end{table}

\subsection{Evidence for the energy transfer process: spectrally--resolved PL dynamics influenced by an external magnetic field}

\begin{figure}
\centering
\includegraphics[width=\linewidth]{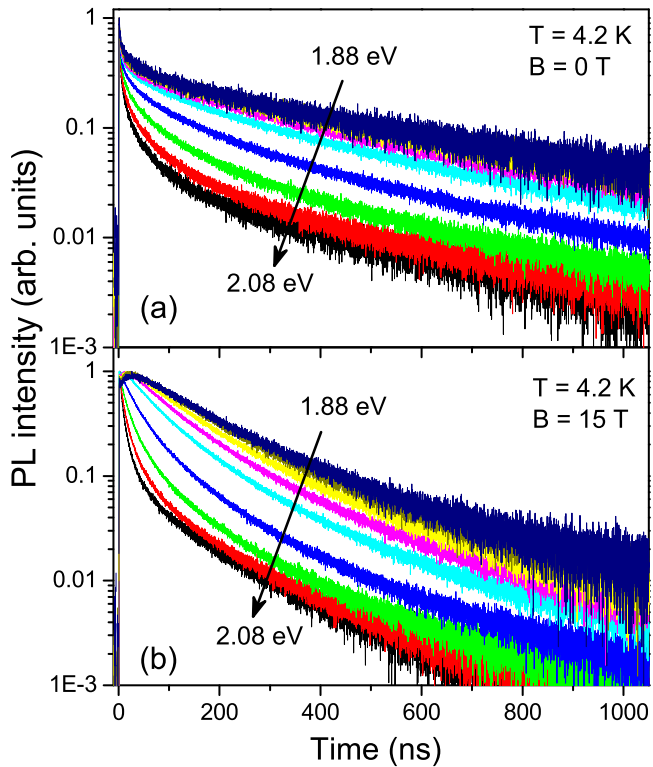}
\caption{Spectrally-resolved PL dynamics of the 3.4~nm CdTe NCs
measured at (a) $B=0$~T and (b) 15~T. The detection energy is varied
in the range from 1.88 to 2.08~eV. The signals are normalized on
their intensity at zero delay, i.e., right after the excitation
pulse.} \label{fig:decay}
\end{figure}

Figure~\ref{fig:decay}shows the spectrally-resolved PL dynamics of
the 3.4~nm CdTe NCs measured at $T=4.2$~K. At zero magnetic field,
the PL decay shows a multiexponential behavior. At the high energy
side of the emission band, e.g., at 2.08~eV (black line), the decay
starts with fast components of 1 and 4~ns that are followed by an
intermediate component of 30~ns finally turning to a very slow
component with 320~ns decay time. The relative contribution of the
intermediate component decreases for lower spectral energies,
whereas the slow component is spectrally independent. The fast
component can be attributed to bright excitons, namely to their
radiative recombination and scattering to the dark states. The
intermediate component is most likely related to the FRET from
smaller to larger NCs. The slow component is due to the
recombination of dark excitons.

The spectrally-resolved PL dynamics change considerably in an
external magnetic field, compare Figs.~\ref{fig:decay}(a) and
\ref{fig:decay}(b). In a magnetic field the bright and dark exciton
states become mixed, which results in the vanishing of the fast
component and the shortening of the slow one. For example, the slow
component of the PL decay at 2.08~eV (black line) shortens from 320
to 120~ns, when the magnetic field is increased to 15~T. Compared
with the PL dynamics measured at the high energy side of the PL
spectra, the PL decay measured at low energy side shows a
qualitatively different behavior. The PL dynamics at 1.88~eV (blue
line) has a small initial rise and then decays, see
Fig.~\ref{fig:decay}(b). This rise can be explained only by the
energy transfer process in which dark exciton states in larger NCs
are fed by smaller NCs.

\begin{figure}
\centering
\includegraphics[width=\linewidth]{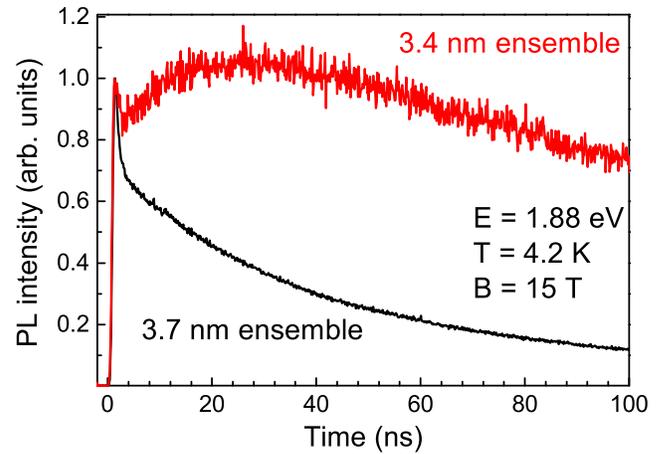}
\caption{Spectrally-resolved PL dynamics of the 3.4-nm and 3.7~nm CdTe NC ensembles measured at the energy of 1.88~eV, see Fig.\ref{fig:pl}(a).} \label{fig:new}
\end{figure}

Further confirmation for the role of the energy transfer in the PL decay dynamics is
shown in Fig.~\ref{fig:new}. Here the PL dynamics from two NC ensembles are measured
in the magnetic field $B=15$~T at the same energy of 1.88~eV, where the low
energy tail of the 3.4~nm ensemble overlaps with the high energy tail of the 3.7~nm
ensemble. Due to the NC size dispersion in the ensembles the specific
 energy corresponds to the specific NC size. While the PL dynamics differ drastically being strongly
 contributed by the energy transfer.

\begin{figure}
\centering
\includegraphics[width=\linewidth]{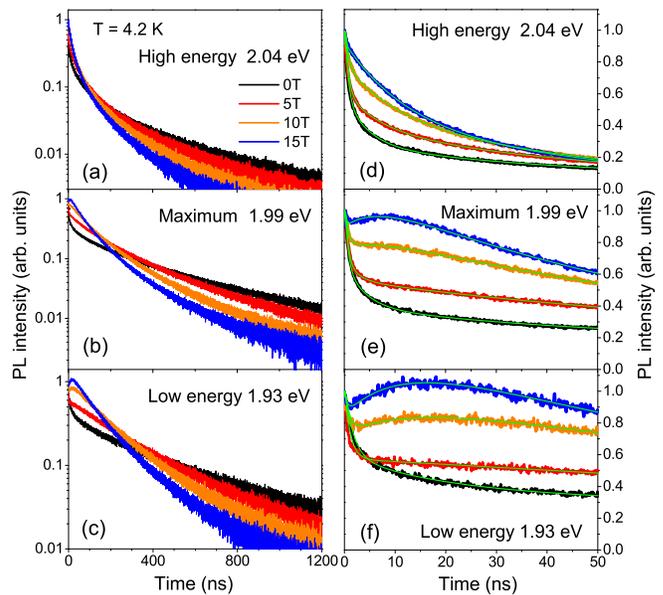}
\caption{(Color online)(a-c) PL dynamics of the 3.4~nm CdTe NCs measured at different
spectral energies (see Fig.~1) and different magnetic fields. Panels
(d-f) detail the PL dynamics during the initial 50~ns. The green
lines show fits to these curves with the function described in
Appendix. } \label{fig:first50ns}
\end{figure}

Figure~\ref{fig:first50ns}(a)-~\ref{fig:first50ns}(c) show the PL dynamics measured at
different magnetic fields for three energies on the 3.4~nm NCs,
compare with Fig.~\ref{fig:pl}. For clarity,
Figs.~\ref{fig:first50ns}(d-f) are close-ups of the initial 50~ns of
the PL dynamics and show their amplitude on a linear scale. The PL
decay at the high energy position shows the behavior typical for colloidal NCs, see Figs.~\ref{fig:first50ns}(a,d). With
increasing magnetic field, the fast component vanishes and the slow
component shortens due to the mixing of bright and dark exciton
states. However, a different behavior is observed at the maximum of PL spectra in Figs.~\ref{fig:first50ns}(b,e). The PL dynamics at
$B=0$~T shows a decay in the time frame $t\in(5~\text{ns},
15~\text{ns})$, which slows down with increasing magnetic field  and
turns into a rise at $B=15$~T. Such unusual behavior is even more
prominent at the low energy position, see
Figs.~\ref{fig:first50ns}(c,f). At $B=15$~T the PL intensity at
$t=12$~ns is larger than the intensity at $t=0$~ns.

\begin{figure}
\centering
\includegraphics[width=\linewidth]{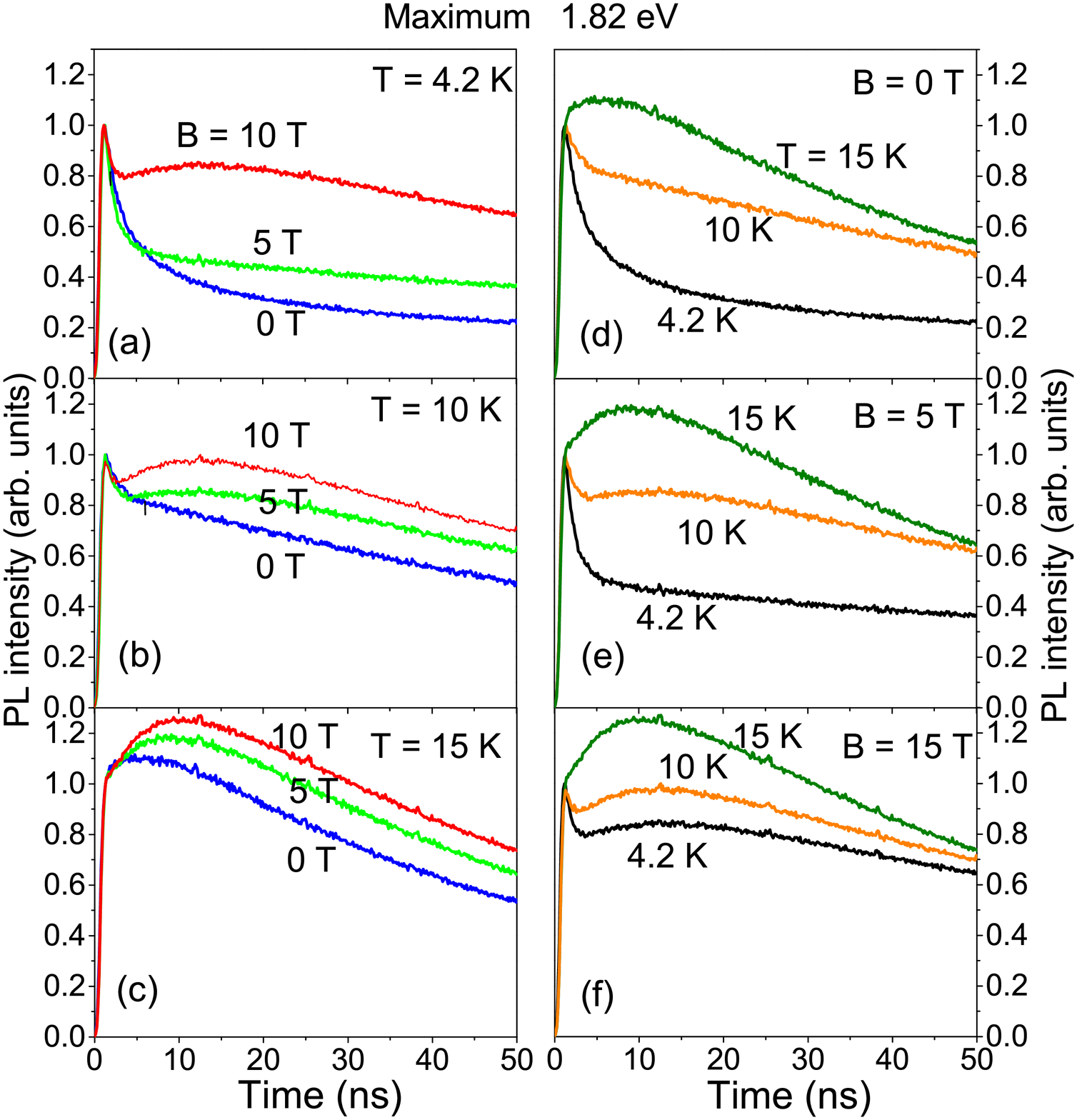}
\caption{PL dynamics of the 3.7~nm CdTe NCs measured at the maximum
of the PL band at 1.82~eV for various magnetic fields and
temperatures.} \label{fig:TvsB}
\end{figure}

A qualitatively very similar behavior was found for the 3.7~nm NCs,
whose PL dynamics during the initial 50~ns measured at the PL
maximum of 1.82~eV is presented in Figure~\ref{fig:TvsB}. In panels
(a-c) the PL dynamics measured at fixed temperature are compared for
different magnetic fields. At $T=4.2$~K the PL dynamics shows a
decay at $B=0$~T. At higher fields the PL intensity in the time
interval $t\in(5~\text{ns}, 15~\text{ns})$ starts to grow, and at
15~T the decay turns into a rise. This behavior is more pronounced
at higher temperatures, see Fig.~\ref{fig:TvsB}(b). At 10~K, the
decay turns to a rise already at $B=5$~T. At even higher temperature
of 15~K the fast decay component within the first 5~ns disappears,
which can be seen in Figs.~\ref{fig:TvsB}(a,b), and only a slow rise
is visible during the initial 10~ns, see Fig.~\ref{fig:TvsB}(c). The
rise time increases for stronger magnetic fields.

Besides the magnetic field, the shape of the PL decay is also
strongly influenced by temperature. As shown in
Fig.~\ref{fig:TvsB}(d), the PL dynamics at 4.2~K and 10~K start with
a fast decay component, while this fast component cannot be seen at
15~K. Instead, the PL dynamics starts with a rise of the PL
intensity. At $B=5$~T the rise of the PL intensity becomes more
prominent, see Fig.~\ref{fig:TvsB}(e). At $B=15$~T the PL dynamics
show this rise during the initial 15~ns at all studied temperatures.
The fast decay component presents at 4.2~K and 10~K, but disappears at
15~K, see Fig.~\ref{fig:TvsB}(f).

\begin{figure}
\centering
\includegraphics[width=\linewidth]{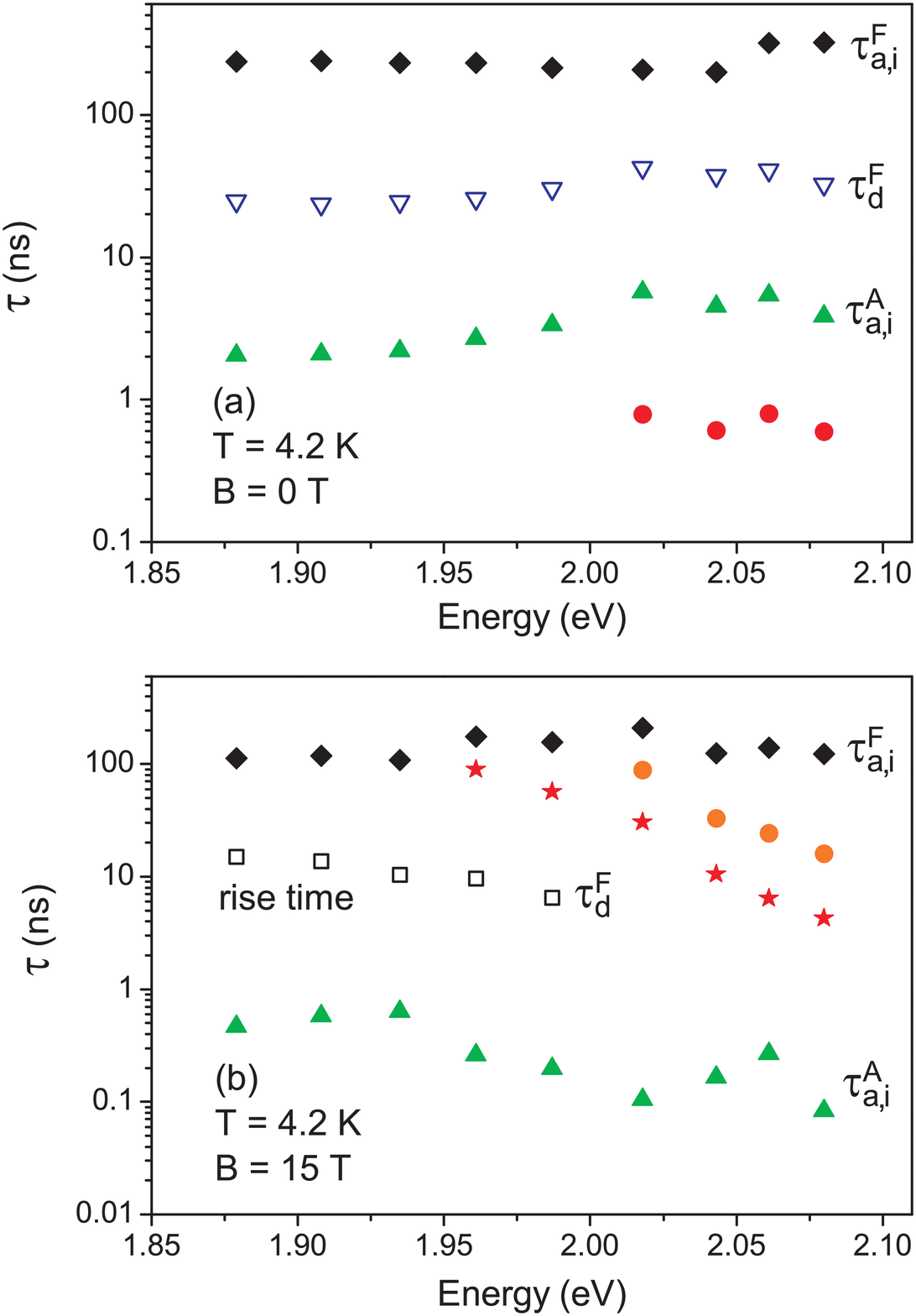}
\caption{Spectral dependence of the PL decay and rise times of the
3.4~nm CdTe NCs for (a) $B=0$~T and (b) 15~T at $T=4.2$~K. The open
squares in (b) show the rise time of the initial rising component
that appears in magnetic field (see, e.g., the blue line in
Fig.~\ref{fig:first50ns}(f)). The PL decay times are obtained by
fitting the spectrally-resolved PL decay shown in
Fig.~\ref{fig:decay} with the multiexponential function of
Eq.~\eqref{EQ:fit} given in Appendix.} \label{fig:TauE}
\end{figure}

Figure~\ref{fig:TauE} shows the spectral dependence of the PL decay
times of the 3.4~nm CdTe NCs for $B=0$~T and 15~T measured at
$T=4.2$~K. The PL decay times were obtained by fitting the
spectrally-resolved PL decays shown in Fig.~\ref{fig:decay} with the
multiexponential function given by Eq.~\eqref{EQ:fit} as described in
Appendix. In Fig.~\ref{fig:TauE}(a) the green triangles correspond
to the fast component $\tau_\text{ai}^\text{A}$ observed at the very beginning
of the PL decay. We relate this decay to the relaxation of excitons
from the bright state to the dark state. The black closed diamonds
correspond to the longest component $\tau_\text{ai}^\text{F}$, related to the
decay of the dark excitons that are not involved in the energy
transfer process as donors. The blue open triangles correspond to
the component $\tau_\text{d}^\text{F}$ related to the dark excitons involved in
the energy transfer process as donors having, therefore, a shortened
lifetime. The amplitude of this component is small at the low energy
side of the PL spectra, where the FRET does not contribute to the PL
dynamics, and increases towards the high energy side of the PL
spectra. The fast component marked by red circles, which appears
only at the high energy side of the PL spectra, is most likely
related to the energy transfer from the bright exciton states. One
can see, that at $B=0$~T the characteristic times of all these
components show no apparent spectral dependence, while their
relative amplitudes are spectrally dependent.

For the PL dynamics at a magnetic field of 15~T the characteristic
times are collected in Fig.~\ref{fig:TauE}(b). Here the open squares
correspond to the rising component induced by the energy transfer.
This component is visible only at the low energy side of the PL
spectra. Compared with the zero-field case, two new components (the red
stars and the orange circles) appear at the high energy side of the
PL spectra, indicating that new energy transfer paths are activated
by the magnetic field or additional NCs become involved in the FRET.
On the basis of these characteristic decay times, we developed a
theoretical model describing the observed experimental results.
Description of the model and presentation of the simulation results
will be done in the following sections.

\begin{figure}
\centering
\includegraphics[width=\linewidth]{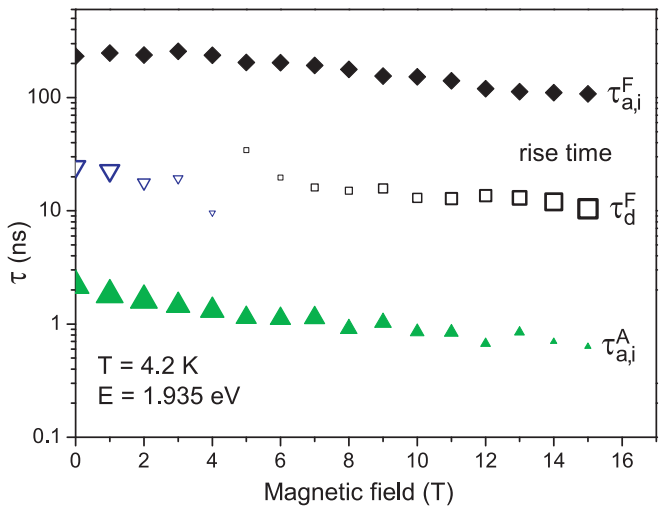}
\caption{Magnetic field dependence of the PL decay and rise times of
the 3.4~nm CdTe NCs measured at 1.93~eV. The size of symbols (except
for the closed diamonds) is proportional to the amplitude of the
corresponding component.} \label{fig:TauB}
\end{figure}

Figure~\ref{fig:TauB} shows the magnetic field dependence of the
characteristic times for the 3.4~nm CdTe NCs measured at the low
energy side of the PL spectra at 1.93~eV for $T=4.2$~K. The shortest
component (green triangles) is related to the bright-dark relaxation
time $\tau_\text{a,i}^\text{A}$. The longest component $\tau_\text{a,i}^\text{F}$
(black closed diamonds) corresponding to the dark exciton decay
shortens by a factor of $\sim2$ with  magnetic field increasing from
0 to 15~T. The component represented by the blue triangles has the
same origin as that in Fig.~\ref{fig:TauE}(a), and is related to
FRET. Its amplitude vanishes with increasing magnetic field. The
open squares correspond to the component with negative amplitude in
the fitting functions (see Appendix), that describes the rise in
the PL dynamics of the acceptor NCs induced by the energy transfer.
This rise time decreases from 34~ns at 0~T down to 10~ns at 15~T.

%%%%%%%%%%%%%%%%%%%%%%%%%%%%%%%%%%%%%%%%%%%%%%%%%%%%%%%%%%%%%%%%%%%%%%%%%%%%

\section{Theoretical considerations}
\label{SecIV}

In this section we present the theoretical model, that describes the
PL intensity of an ensemble of NCs taking into account the energy
transfer process between them. We start from the general assumptions
about the properties of the NC ensemble  and consider the donor and
the acceptor NCs participating in the energy transfer process as well as
the independent NCs (not participating in the FRET process). We discuss
the possible recombination, relaxation and energy transfer pathways
for excitons and write down a system of rate equations for the
exciton populations in donor, acceptor and independent NCs.

\subsection{General assumptions}
\label{SecIVGR}

\begin{figure}[h!]
\centering
\includegraphics[width=8cm]{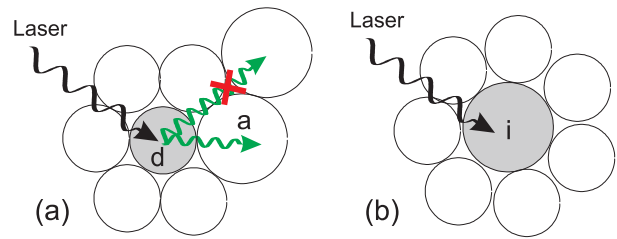}
\caption{Scheme of possible processes initiated by optical
excitation in a NC ensemble: (a) energy transfer between the nearest
donor and acceptor NCs, and (b) independent NCs for which no energy
transfer occurs. } \label{fig:scheme}
\end{figure}

For the sake of clarity, let us consider an ensemble of
closely--packed prolate CdTe NCs with the ground exciton state split
into the bright (optically-allowed A exciton with spin projection
$\pm 1$ on the anisotropy axis) and the dark (optically-forbidden F
exciton with spin projection $\pm 2$) exciton states. Since the
probability of the F\"orster energy transfer decreases proportional
to $R_{\rm da}^\text{-6}$, where $R_{\rm da}$ is the distance
between the centers of the donor and acceptor NCs, it is the largest for the
nearest neighbor NC pairs and already much smaller for the next
nearest neighbors. Thus, in the model we consider the energy
transfer process only between the nearest neighbors. In this case
the energy transfer rate is determined by the magnitude of exciton
dipole moments in the donor and acceptor NCs and by the overlap
integral between the donor emission and acceptor absorption spectra,
while the average distance can be estimated as the mean size of the
NCs: $R_{\rm da}\approx d$. We consider the low excitation regime,
which is defined by assuming that the portion of initially excited
NCs at each energy ($N_\text{0}(E)$) within the NC ensemble is
negligible. This allows us to consider the energy transfer only to
initially unexcited NCs. In turn, initially excited NCs with the
ground state exciton energy $E$ may act as donors (with probability
$f_\text{d}(E)$) and transfer their excitation to the nearest
acceptor NCs (see Fig.~\ref{fig:scheme}(a)), or as independent NCs
which do not participate in the energy transfer process (see
Fig.~\ref{fig:scheme}(b)). Thus, the total number of initially
excited donor and independent NCs is
$N_\text{d}^\text{0}(E)=N_\text{0}(E)f_\text{d}(E)$ and
$N_\text{i}^\text{0}(E)=N_\text{0}(E)(1-f_\text{d}(E))$,
respectively. The properties of the energy-dependent probability
function $f_\text{d}(E)$ will be discussed later. Also, we neglect
all cascade processes, so that one NC cannot act as donor and
acceptor simultaneously.

After these general remarks on the assumptions related to the ensemble of NCs, let us consider the energy transfer mechanism
between the donor and acceptor NCs in more detail. Let us consider a
NC with the ground state exciton energy $E_\text{a}$ at the low
energy side of the PL band arising from the acceptors. A smaller NC
with higher ground state exciton energy
$E_\text{d}=E_\text{a}+E_\text{da}$ may play the role of a donor for
the chosen acceptor if there is a nonzero overlap between the donor
emission spectrum and the acceptor absorption spectrum. We assume
for simplicity that for each $E_\text{d}$ there is only one excited
level $E_\text{a}+E_\text{da}$ in the nearest acceptor NCs and
consider the energy difference  $E_\text{da}$ as a parameter in our
model.

\begin{figure}
\centering
\includegraphics[width=\linewidth]{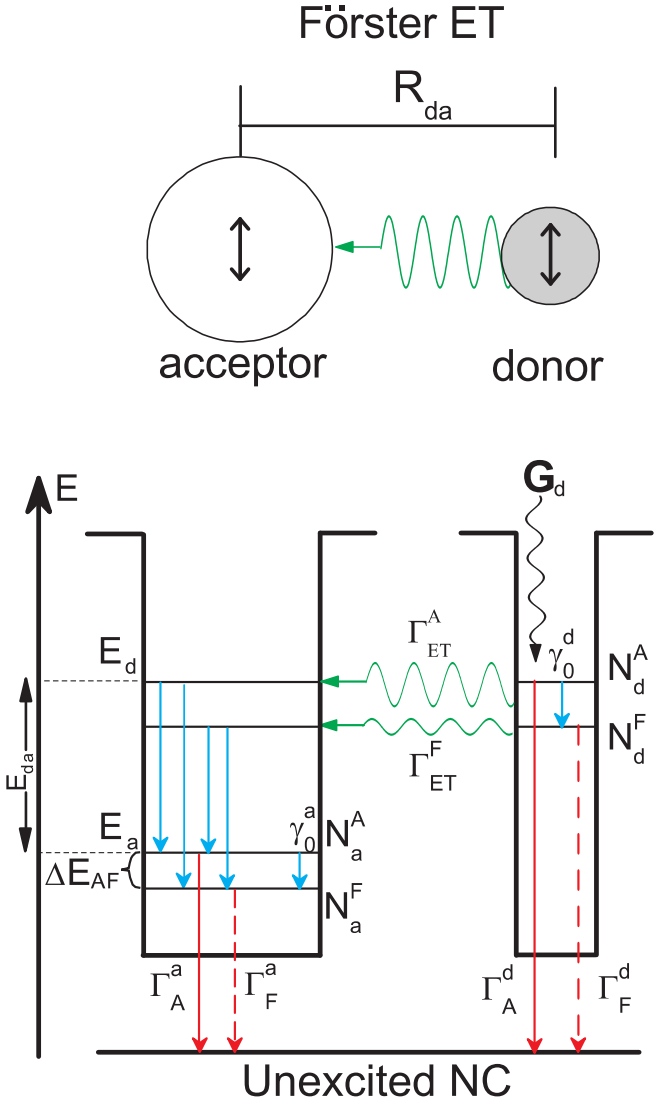}
\caption{Schematic of the energy transfer process between a donor NC
with energy $E_\text{d}$ and an acceptor NC with energy
$E_\text{a}$. The energy transfer is shown by the green arrows, the
relaxation processes by the blue arrows, the  recombination
pathes by the red arrows, and the laser pumping by the black arrow.}
\label{fig:ETscheme}
\end{figure}

In case of initial nonresonant excitation with the rate $G$ using
photon energies well above the ground exciton state energy the
relaxation of the hot excitons to the ground state is a fast process
assisted by optical phonons and the interaction with the
surface.\cite{Klimov2000} After relaxation of the exciton to the
ground state it can be resonantly transferred to another (acceptor)
NC. The energy transfer process between the donor NC with ground
state exciton energy $E_\text{d}$ and the acceptor NC with ground
state exciton energy $E_\text{a}$ is shown schematically in
Fig.~\ref{fig:ETscheme}. We assume that the bright
(optically-allowed A) and dark (optically-forbidden F) exciton
states have approximately the same energy splitting $\Delta E_{\rm
AF}$ in the donor and acceptor NCs. The energy transfer process
consists of two steps. Initially the resonant transfer from the
ground energy state of the donor NC to the first excited energy
state of the acceptor NC takes place. We assume that the bright and
dark ground exciton states in the donor NC are in resonance with the
first excited bright and dark states in the acceptor NC,
respectively. After the energy transfer, fast energy relaxation
from the excited to the ground state of the acceptor NC occurs. This
relaxation process may occur in two different ways, either spin
conserving and thus allowing relaxation only from the bright to the
bright states or from the dark to the dark state or spin non
conserving. For the sake of clarity, we assume here the energy
difference $E_\text{da}$ to be large enough for the fast non
conserving relaxation processes, similar to those taking place after
the initial excitation of all NCs. In this case the relaxation from
the bright and dark excited states occurs with equal probability to
the ground bright and dark states without any "memory" effects.
However, the model can be easily modified for the case of spin
conserving relaxation in the acceptor NCs.

\subsection{Rate equations for the bright and dark exciton populations}
\label{SecIVRE}

To be able to describe the experimental data with our model we
express the PL intensity in terms of exciton populations in the NCs.
The total population of excitons $N(E,t)$ with a given ground state
exciton energy $E$ can be written as the sum of exciton populations
in the donor NCs, $N_{\rm d}$, the independent NCs, $N_{\rm i}$, and
the acceptor NCs, $N_{\rm a}$, NCs. In turn, the population of excitons
in each type of NCs consists of the populations of excitons in the
bright, $N_{\rm d,i,a}^{\rm A}$, and dark,  $N_{\rm d,i,a}^{\rm F}$,
states:
\begin{eqnarray}
N(E,t)=N_\text{i}(E,t)+N_\text{d}(E,t)+N_\text{a}(E,t), \, \\
N_\text{d,i,a}(E,t)=N_\text{d,i,a}^\text{A}(E,t)+N_\text{d,i,a}^\text{F}(E,t). \,
\end{eqnarray}

Using the populations of the bright and dark excitons in each type of NCs, one
obtains for the PL intensity of a NC ensemble at a given exciton energy $E$
\begin{eqnarray}
I(E,t) =  I_\text{i}(E,t)+I_\text{d}(E,t)+I_\text{a}(E,t),  \, \label{inten} \\
I_\text{d,i,a}(E,t) = \Gamma_\text{A}^\text{rad}(E)N_\text{d,i,a}^\text{A}(E,t) +\Gamma_\text{F}^\text{rad}(E)N_\text{d,i,a}^\text{F}(E,t) \, .
\end{eqnarray}
Here $\Gamma_\text{A,F}^\text{rad}$ are the radiative recombination rates.
Thus, knowing the time evolution of the populations $N_\text{d,i,a}^\text{A,F}(E,t)$, one can fully describe the time evolution of the PL intensity from a NC ensemble.

The pair of the bright-dark exciton rate equations for the donor and
the independent NCs, can according to Fig. \ref{fig:ETscheme}, be
written as:
\begin{eqnarray}
\frac{dN_\text{d,i}^\text{A}(E,t)}{dt} &=& -  N_\text{d,i}^\text{A}(\Gamma_\text{A}+\gamma_0+\gamma_\text{th}+ \Gamma_\text{ET}^\text{A})  \nonumber  \\
&+& N_\text{d,i}^\text{F}\gamma_\text{th}+\frac{1}{2}G_\text{d}(E,t)\, , \\
\frac{dN_\text{d,i}^\text{F}(t)}{dt} &=& - N_\text{d,i}^\text{F}(t)(\Gamma_\text{F}+\gamma_\text{th}+\Gamma_\text{ET}^\text{F})  \nonumber \\
&+& N_\text{d,i}^\text{A}(t)(\gamma_0+ \gamma_\text{th})+\frac{1}{2}G_\text{i}(E,t)  \,. \label{donor}
\end{eqnarray}
Here the recombination rates $\Gamma_\text{A,F}=\Gamma_\text{A,F}^{\rm d}$
(which may include both radiative and on nonradiative paths other than the
energy transfer), as well as the relaxation rates $\gamma_0=\gamma_0^{\rm d}$ and
$\gamma_\text{th}=\gamma_\text{th}^{\rm d}$ (thermally induced relaxation given for the
case of a one-phonon process by $\gamma_\text{th} = \gamma_0/[\exp(\Delta E_{\rm AF}/k_{\rm
B}T)-1]$) should be taken at the exciton energy $E_\text{d}$. The energy transfer rates
are denoted as $\Gamma_\text{ET}^\text{A}$ and $\Gamma_\text{ET}^\text{F}$ for the bright
and dark exciton states in the donor NCs, respectively, and have to be set to zero in the
independent NCs. The rates of pumping of the bright and dark exciton states are denoted
as $\frac{1}{2}G_\text{d,i}(E,t)$ and are equal because of the spin non conserving
relaxation of the hot excitons, as mentioned previously.

The rate equations for the acceptor NCs, according to Fig.
\ref{fig:ETscheme}, can be written as:
\begin{eqnarray}
\frac{dN_\text{a}^\text{A}(E,t)}{dt} &=& - N_\text{a}^\text{A}(\Gamma_\text{A}+\gamma_0+\gamma_\text{th}) + N_\text{a}^\text{F} \gamma_\text{th}  \nonumber \\
&+& \frac{1}{2} (N_\text{d}^\text{A}\Gamma_\text{ET}^\text{A}+ N_\text{d}^\text{F}\Gamma_\text{ET}^\text{F} )\, ,  \\
\frac{dN_\text{a}^\text{F}(E,t)}{dt} &=& - N_\text{a}^\text{F}((\Gamma_\text{F}+\gamma_\text{th}) + N_\text{a}^\text{A}(\gamma_0 + \gamma_\text{th})  \nonumber \\ &+& \frac{1}{2}(N_\text{d}^\text{A} \Gamma_\text{ET}^\text{A}+ N_\text{d}^\text{F}\Gamma_\text{ET}^\text{F} ) \, .\label{acceptor}
\end{eqnarray}
Here the population $N_\text{a}$ should be considered at the exciton
energy $E_\text{a}$ (as well as the recombination rates
$\Gamma_\text{A,F}=\Gamma_\text{A,F}^{\rm a}$  and the relaxation
rates $\gamma_0=\gamma_0^{\rm a}$ as well as
$\gamma_\text{th}=\gamma_\text{th}^{\rm a}$ rates), while the
population $N_\text{d}$ has to be taken at the energy
$E_\text{a}+E_\text{da}$. Since we consider the low excitation
regime, a pumping term is not included in the rate equations for the
acceptor NCs.

\section{Modeling of the experimental data}
\label{SecV}

\subsection{Determination of decay times for donor, acceptor and independent NCs}
\label{SecVD}

The system of rate equations described in subsection \ref{SecIVRE}
can be solved numerically or analytically. The analytical solution
can be simplified in the low temperature limit, assuming
$\gamma_\text{th}^{\text{a(d)}} \approx 0$. Additionally, we consider the
following hierarchy of rates
$\gamma_0\gg\Gamma_\text{A}\gg\Gamma_\text{F}$ and also assume that
all rates do not depend on energy. This allows us to obtain simple
approximate expressions for the time evolution of the populations
$N_\text{d,i,a}^\text{A,F}(E,t)$, which depend on the energy $E$
only via the energy dependent initial conditions
$N_\text{d,i,a}^\text{A,F}(E,t=0)$. These solutions can be written
for donor NCs and independent NCs as:
   \begin{eqnarray}
 && N_\text{d,i}^\text{A}(E,t)= N_\text{d,i}^\text{A}(E,t=0) \exp(-t/\tau_\text{d,i}^\text{A}) \, , \\
 && N_\text{d,i}^\text{F}(E,t) \approx
  - N_\text{d,i}^\text{A}(E,t)+
  N_\text{d,i}^0(E)\exp(-t/\tau_\text{d,i}^\text{F}) \, \label{DT}
 \end{eqnarray}
and for the acceptor NCs as
  \begin{eqnarray}
&&N_\text{a}^\text{A}(E,t)
\approx   N_\text{d}^\text{A}(E+E_\text{da},t=0)  \times \nonumber \\
 &&
\left[ \exp(-t/\tau_\text{a}^\text{A})-\exp(-t/\tau_\text{d}^\text{A}) \right] \, ,   \\
&& N_\text{a}^\text{F}(E,t)= - N_\text{a}^\text{A}(E,t)+ \nonumber  \\
 && N_\text{d}^0(E+E_\text{da}) \left[ \exp(-t/\tau_\text{a}^\text{F}) -  \exp(-t/\tau_\text{d}^\text{F}) \right] \, .\label{AT}
 \end{eqnarray}
Here we use $N_\text{d,i}^0(E)=
N_\text{d,i}^\text{A}(E,t=0)+N_\text{d,i}^\text{F}(E,t=0)$ and
$N_\text{a}^0(E)=
N_\text{a}^\text{A}(E,t=0)+N_\text{a}^\text{F}(E,t=0)=0$. The
characteristic times are
\begin{eqnarray}
&&\frac{1}{\tau_\text{d}^\text{A}}= \Gamma_\text{A}+ \gamma_0+\Gamma_\text{ET}^\text{A} \, , \\
&&\frac{1}{\tau_\text{d}^\text{F}}= \Gamma_\text{F}+\Gamma_\text{ET}^\text{F} \, \\
&&\frac{1}{\tau_\text{a,i}^\text{A}}= \Gamma_\text{A}+ \gamma_0 \, , \\
&&\frac{1}{\tau_\text{a,i}^\text{F}}= \Gamma_\text{F} \, .
\end{eqnarray}
The equations allow us to reveal the decay times for each kind of
NCs. By comparing these times with the times extracted from the
multi-exponential fit of the experimental decay curves we evaluate
the rates $\Gamma_\text{F,A}$, $\Gamma_\text{ET}^\text{F,A}$ and
$\gamma_0$. We assume that the recombination rate of the dark
exciton $\Gamma_\text{F}$ and its magnetic field dependence can be
directly associated with the longest decay component
$\Gamma_\text{F}=1/\tau_\text{a,i}^\text{F}$ obtained by fitting the
experimental decay curves and shown by the black diamonds in
Figs.~\ref{fig:TauE} and \ref{fig:TauB}. One sees that this rate
indeed very weakly depends on the energy but increases nearly by a
factor of two with increasing magnetic field up to 15~T. This
increase is caused by the fact, that despite of the cubic symmetry
of the CdTe crystal lattice, the NCs possess an anisotropic axis
related to their nonspherical shape.\cite{Liu2013} As a result, a
magnetic field having nonzero projection on the anisotropy axis
mixes the bright and dark exciton states similar to the well known
situation in hexagonal CdSe NCs.\cite{Efros1996}

Furthermore, the bright exciton recombination rate
$\Gamma_\text{A}=0.1$ ns$^{-1}$ and the value of the bright-dark
splitting $\Delta E_\text{AF}=2.2$ meV are obtained from the
temperature dependence of the PL decay (not shown
here).\cite{Labeau2003}   We  determine the relaxation rate
$\gamma_0$ and its magnetic field dependence from the bright exciton
lifetime $\tau_\text{a,i}^\text{A}$ at $E_\text{a}=1.93$~eV (the
green triangles in Fig.~\ref{fig:TauB}) as
$\gamma_0=1/\tau_\text{a,i}^\text{A}-\Gamma_\text{A}$. The energy
transfer rate $\Gamma_\text{ET}^\text{F}$ can be found from the
difference of the dark exciton lifetimes
$\Gamma_\text{ET}^\text{F}=1/\tau_\text{d}^\text{F}-1/\tau_\text{a,i}^\text{F}$.
However, attempts to estimate this rate from the decay components at
the high energy (donor) side of the spectrum are complicated by the
fact that there are more components than just one in high magnetic
field and their rates depend on the spectral position, see
Fig.~\ref{fig:TauE}(b). For this reason we use the rise rates (the
open squares in Figs.~\ref{fig:TauB} and \ref{fig:TauE}(b)) observed
at the low energy side of the spectrum at $B\geq4$~T in order to
estimate $\tau_\text{d}^\text{F}$ averaged over the spectral
position and to obtain the magnetic field dependence of the
$\Gamma_\text{ET}^\text{F}$.  The determination of the energy
transfer rate $\Gamma_\text{ET}^\text{A}$ from the bright exciton is
more difficult because of its short lifetime. We estimate that the
energy transfer from the bright exciton state is faster than the
relaxation to the dark exciton state in the donor NCs. The
parameters determined from the analysis of the multi-exponential fit
for magnetic fields of $B=0$, 5, 10 and 15~T are summarized in
Table~\ref{tab:model2}.

\begin{table}
\centering
\begin{tabular}{|c|c|c|}
\hline
\phantom{(T)}&\phantom{(ns$^{-1}$),(ns$^{-1}$),(ns$^{-1}$) }& \phantom{(ns$^{-1}$),(ns$^{-1}$),(ns$^{-1}$) }\\
\phantom{(T)}& multi-exponential fit&  rate equations \\
\hline
\end{tabular}
\begin{tabular}{|c|c|c|c|c|c|c|}
\hline
$B$ & $\gamma_\text{0}$ & $\Gamma_\text{F}$  & $\Gamma_\text{ET}^\text{F}$ & $\Gamma_\text{F}$  & $\Gamma_\text{ET}^\text{F}$  & $w$ \\
(T) & (ns$^{-1}$) & (ns$^{-1}$) & (ns$^{-1}$) & (ns$^{-1}$) &  (ns$^{-1}$) & \phantom{(ns$^{-1}$)} \\ \hline
0 & 0.45 & 0.004 &   & 0.0035 & 0.055 & 0.23\\ \hline
5 & 0.72 & 0.005 & 0.023 &  0.0040  & 0.055 & 0.17\\ \hline
10 & 1.0 & 0.007 & 0.069  & 0.0065 & 0.070 & 0.13\\ \hline
15 & 1.3  & 0.009 & 0.090  & 0.0085 & 0.090 & 0.11\\ \hline
\end{tabular}
\caption{Magnetic--field dependent parameters determined from the
analysis of the multi-exponential fit and from the simulation of the
PL dynamics with the rate equations. $\Gamma_\text{ET}^\text{F}$ is
the energy transfer rate from the dark exciton state,
$\Gamma_\text{F}$ is the  recombination rate of the dark exciton
state, and  $\gamma_\text{0}$ is the relaxation rate from the bright
to the dark exciton states at $T=0$~K (taken to be the same for all
donor and acceptor NCs). The parameter $w$ is the fraction of bright
excitons in independent NCs at time $t=0$.} \label{tab:model2}
\end{table}

\subsection{Shift of the CW PL spectrum due to the nonradiative energy transfer}
\label{SecVCW}

The PL spectrum measured at time $t=0$ can be approximated by a Gaussian form with a peak
energy at $E=E_0$:
\begin{eqnarray}
I(E,t=0) = \frac{I}{\sigma \sqrt{2\pi}} \exp \left[-\frac{(E - E_0)^2}{2\sigma^2}  \right],
\label{EQ:shape}
\end{eqnarray}
where $I$ is the total spectrally-integrated intensity at $t=0$ and
$\sigma$ corresponds to the linewidth at half maximum according to
$2\sigma\sqrt{2\text{ln}2}$. Neglecting the spectral dispersion of
the recombination rates of the bright and the dark excitons
$\Gamma_\text{A}$ and $\Gamma_\text{F}$ as well as their relative
populations at $t=0$ we assume that the shape of the time-resolved
PL spectrum at $t=0$ reflects the initial energy dispersion of the
excitons populations: $I(E,t=0) \propto N^0(E)$.

To simulate the temporal shift of the CW PL peak position compared
to the peak position at time $t=0$ we solve the system of rate
equations in the steady state regime. Considering the low
temperature limit and constant generation rates $G_\text{d,i} =
GN_{\rm d,i}^0(E)$ we obtain the following solutions for the
populations in the different kinds of NCs:
\begin{eqnarray}
N_\text{d,i}^\text{A}(E)&=& \frac{1}{2}G\tau_\text{d,i}^\text{A}N_\text{d,i}^\text{0}(E) \, , \label{eq:20} \\
N_\text{d,i}^\text{F}(E)&=&\frac{\tau_\text{d,i}^\text{F}}{2}\left[GN_\text{d,i}^\text{0}(E)+\gamma_0N_\text{d,i}^\text{A}(E) \right]\, , \label{eq:21}\\
N_\text{a}^\text{A}(E)&=&\frac{\tau_\text{a}^\text{A}}{2}\left[\Gamma_\text{ET}^\text{A}N_\text{d}^\text{A}(E_\text{d})
+\Gamma_\text{ET}^\text{F}N_\text{d}^\text{F}(E_\text{d})\right]\, , \label{eq:22}\\
N_\text{a}^\text{F}(E)&\approx&2N_\text{a}^\text{A}(E)\frac{\tau_\text{a}^\text{F}}{\tau_\text{a}^\text{A}} \,. \label{eq:23}
 \end{eqnarray}
where $E_{\rm d}=E+E_\text{da}$. The final equation describing the
CW spectrum is rather cumbersome but can be simplified by taking
into account the ratio of bright and dark exciton lifetimes. From
Eqs.~(\ref{eq:20}-\ref{eq:23}) one can see that the populations of
the bright and dark excitons are proportional to their
characteristic lifetimes. As the lifetime of the dark excitons is
two orders of magnitude longer than the bright exciton lifetime we
can neglect the contribution of the bright excitons. In this case a
simple equation for the shift of the CW spectrum can be obtained:
\begin{eqnarray}
&&I_\text{PL}^{\text{CW}}(E)=I(E,t=0)\left[ 1-K_\text{ET} T_{\rm d}(E) \right]\, , \label{eq26} \\
&&T_{\rm d}(E)=f_\text{d}(E)-f_\text{d}(E+E_\text{da})\frac{N_0(E+E_\text{da})}{N_0(E)} \, , \label{eq27}
\end{eqnarray}
where
\begin{equation}
K_{\rm ET}=\frac{\Gamma_\text{ET}^\text{F}}{\Gamma_\text{ET}^\text{F}+\Gamma_\text{F}}
\label{ket}
\end{equation}
describes the efficiency of the energy transfer process from the dark exciton state. The
physical meaning of the transfer function $T_{\rm d}$ is clearly seen: the excitation is
transferred from the spectral region $T_{\rm d}(E)>0$ to the region where $T_{\rm
d}(E)<0$.

We choose for the function $f_\text{d}(E)$ a Gaussian form with the
peak at $E_0+E_\text{da}$. To keep the energy distance $E_\text{da}$
at the same (average) value over the spectrum, we allow the
dispersion $\sigma_\text{d}$ of the $f_\text{d}(E)$ to be larger
than $\sigma$. Comparison of the CW spectrum with the time-resolved
spectrum at $t=0$ allows us to determine the dispersion
$\sigma_\text{d}=90 $ meV and the average value of $E_\text{da}=106
$ meV.  The comparison of the exact solution with the solution given
by Eq. (\ref{eq26}) shows no difference. Physical meaning is that the fast relaxation of excitons from the bright state
with the rate $\gamma_\text{0}$ results in the accumulation of excitons
in the dark state. In case of the CW excitation this is equivalent to the
direct generation of dark excitons only.

\begin{figure}
\centering
\includegraphics[width=8 cm]{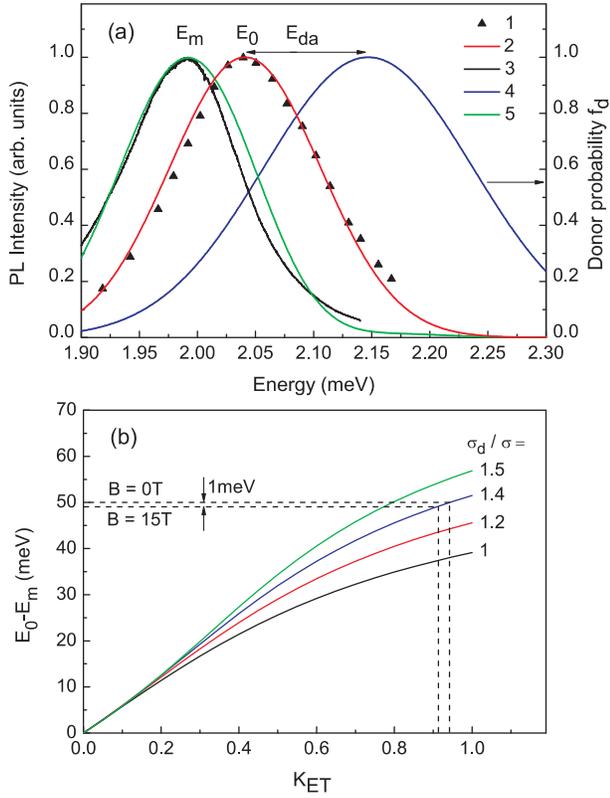}
\caption{(a) Modeling of the PL spectra: (1) Experimentally measured
time-resolved PL spectrum of the 3.4 nm NCs at $t=0$; (2) Gaussian-shaped
time-resolved PL spectrum $I(E,t=0)$ with the peak at $E_0$;
(3)Experimentally measured CW PL spectrum with the peak at $E_m$;
(4) Donor probability function $f_{\rm d}$; (5) Modeled CW PL
spectrum given by the solution of the rate equations in the
steady-state regime. (b) Modeled dependence of the PL shift maximum
$E_0-E_m$ on the energy transfer efficiency $K_{\text{ET}}$.}
\label{shift}
\end{figure}

In the absence of the bright excitons in the donor NCs the shift of the
PL peak in the CW regime is determined by the energy transfer
efficiency $K_{\rm ET}$ from the dark excitons. For modeling the
spectral shift in Fig. \ref{shift}(a) we used $K_{\rm ET} \approx
0.9$ determined for $B=15$ T with the rates given in Table
\ref{tab:model2}. Figure \ref{shift}(b) shows the dependencies of
the PL maximum shift $E_0-E_m$ on the ET efficiency $K_{\rm ET}$ for
different values of the ratio $\sigma_{\rm d}/\sigma$. One sees,
that in the range of large $K_{\rm ET}$ close to unity, the PL
maximum shift $E_0-E_m$ changes insignificantly. We remind also,
that the experimentally measured unpolarized CW spectra in different
magnetic fields differ insignificantly as well (compare, for
example, the CW spectra for $B=0$ (the black line) and $B=15$~T (the
red line) in Fig.~\ref{fig:pl}(a)).  The nonlinear dependence of
$E_0-E_m$ on $K_{\rm ET}$ can be obtained from Eq.~(\ref{eq26})  as
\begin{eqnarray}
E_0-E_m= \Delta (E_m) K_{\rm ET} \, , \\  \Delta
(E)=\frac{\sigma^2}{N_0(E)}\frac{\partial [N_0(E)T_{\rm
d}(E)]}{\partial E} \, .\label{DENT}
\end{eqnarray}

Note, that the recombination rate $\Gamma_\text{F}$ in Eq.~(\ref{ket}) may
include also the nonradiative decay path other than the energy transfer. The presence of
fast nonradiative recombination could prevent the observation of the  spectral shift of the PL line caused by the energy transfer due to decrease of the $K_\text{ET}$ value. It is not the case for the studied samples, where $K_\text{ET}$ very close to unity is evaluated. To simplify
the following modeling of the recombination dynamics we neglect the nonradiative
mechanisms other than ET and assume hereafter the recombination rates to be purely
radiative: $\Gamma_\text{A,F}=\Gamma_\text{A,F}^\text{rad}$.

\subsection{Spectral dependence of the recombination dynamics }
\label{SecVDA}

To simulate the recombination dynamics of the NC ensemble, we combine
the solution of the system of rate equations for a donor-acceptor
NC pair in the transient regime with the determined
probability function $f_{\rm d}(E)$ for the energy transfer.
However,  the initial conditions at $t=0$ for the bright excitons,
$N_\text{d,i}^\text{A}(E,t=0)$, and the dark excitons,
$N_\text{d,i}^\text{F}(E,t=0)$, should be determined first. The
relative populations of the bright and the dark excitons,
$N_\text{d,i}^\text{A}(E,t=0)$ and $N_\text{d,i}^\text{F}(E,t=0)$,
and their corresponding contributions to the PL may depend on the
conditions of excitation and detection. For photoexcitation with
short laser pulses, the relaxation, recombination and energy
transfer processes start simultaneously. For our simulations at all energies, we choose as the initial time $t = 0$ the time when the
initial growth of the PL intensity after the pumping pulse turns
into a decay. Even if the exciton relaxation to the ground state
occurs with an equal probability to the dark and bright states, their
populations at $t=0$ might be not equal. The reason is that the
pulse duration (see Appendix) is comparable with the relaxation time
between the bright and dark exciton states and with the energy
transfer rate from the bright excitons. Hence, we introduce the
initial condition for the bright exciton population in the
independent NCs as $N_\text{i}^\text{A}(E,t=0)=wN_\text{i}^0(E)$ and
consider $w$ as an energy independent parameter. The additional
growth of the PL intensity after $t=0$  is observed at the low
energy side of the spectrum at $4.2$ K only, with the rate
corresponding to the energy transfer from the dark exciton (see
Fig.~\ref{fig:first50ns}(e,f), Fig.~\ref{fig:TauE}(b) and
Fig.~\ref{fig:TauB}). For this reason  we neglect in our simulation
the energy transfer from the bright exciton states and use the
initial condition for the donor NCs $N_\text{d}^\text{A}(E,t=0)=0$.

\begin{figure}[h!]
\centering
\includegraphics[width=\linewidth ]{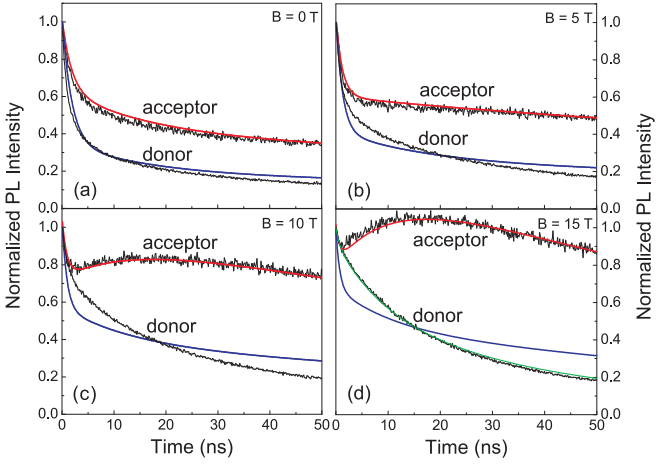}
\caption{Experimental data (the black curves) and calculations of
the PL dynamics of acceptor NCs at $E_\text{a}=1.93$~eV (the red
curves) and donor NCs at $E_\text{a}=2.04$~eV (the blue curves) at
$T=4.2$~K for $B=0$, 5, 10 and 15~T. The calculation results were
achieved from solution of the rate equations Eqs.~\eqref{donor} and
\eqref{acceptor} using the parameters listed in the
Table~\ref{tab:model2}. The green curve for $B=15$~T corresponds to
the solution of the rate equations with an additional non radiative
process (see the description in text).} \label{DApair}
\end{figure}

Using these initial conditions, a consistent modeling of the PL decay
$I(E,t)$ for NCs with emission energies at
$E=E_\text{d}=2.04~\text{eV}$ and $E=E_\text{a}=1.93~\text{eV}$ can
be achieved. This modeling allows us to refine the values of the
rates $\Gamma_\text{F}$ and $\Gamma_\text{ET}^\text{F}$  by
comparing the simulated decay curves with the experimental data
measured at $T=4.2$~K for magnetic fields $B=0-15$~T at the emission
energies $E_\text{a}=1.93$ eV and $E_\text{d}=2.04$ eV, see Figs.~\ref{fig:first50ns}(d) and ~\ref{fig:first50ns}(f). The refined
parameters used in the modeling are listed in
Table~\ref{tab:model2}. The results of the calculations are
presented in Fig.~\ref{DApair}. One sees that a good agreement with
the experimental data is achieved for the PL decay of the NCs at
$E_\text{a}$ (see the red curves). For the PL decay  at $E_\text{d}$
(blue curves) the difference between the simulated and the
experimental decay curves increases with the increasing magnetic field.
Apparently, this difference is caused by neglected additional non
radiative processes (for example, additional energy transfer to NCs
other than those emitting at $E_\text{a}=1.93$~eV), which are
indicated by the additional decay times shown in
Fig.~\ref{fig:TauE}(b) for the high energy emission of the PL in a
magnetic field of $B=15$~T. Accounting for the additional process
with a decay rate of $0.035$~ns$^{-1}$ corresponding to the orange
circles in Fig.~\ref{fig:TauE}(b) for the NCs at
$E_\text{d}=2.04~\text{eV}$ allows us to simulate the decay curves
(see the green curve in Fig.~\ref{DApair}(d) with better accuracy.

\begin{figure}[h!]
\centering
\includegraphics[width=8 cm]{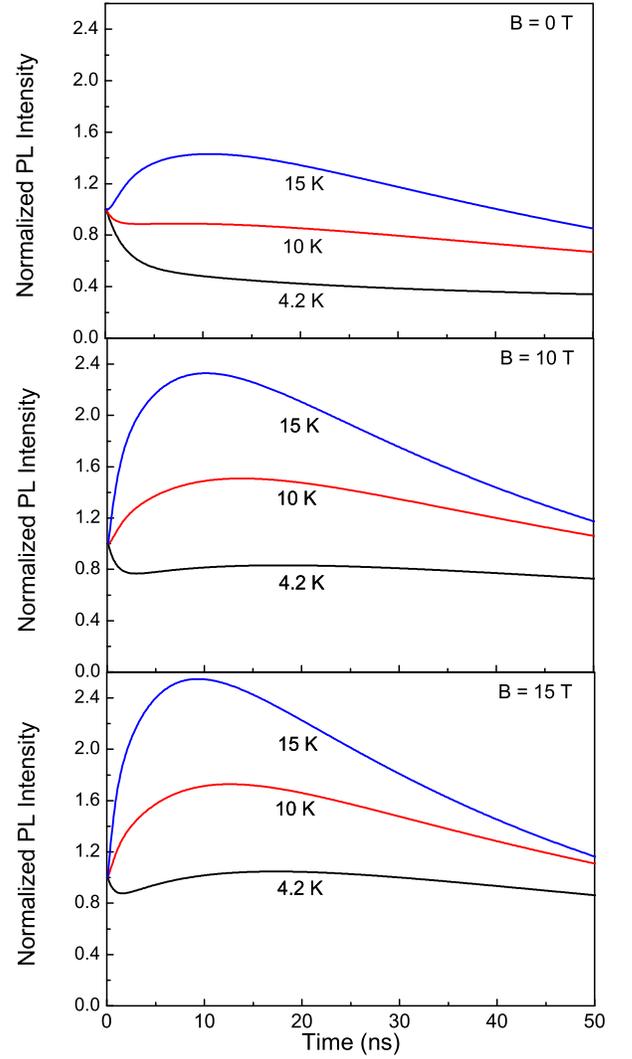}
\caption{ Modeling of the  PL dynamics of the 3.4 nm NCs at energy
$E=E_\text{a}=1.93$~eV for the temperatures $T=4.2$~K (the black
curves), 10~K (the red curves) and 15~K (the blue curves) in
magnetic fields of $B=0$ (a), 10~T (b)  and 15~T (c). The parameters
used for the modeling are given in  text and in Table~\ref{tab:model2}. }
\label{Temperature}
\end{figure}

Using the refined parameters for the recombination and energy
transfer rates we can also simulate the PL dynamics of the acceptor
NCs at different temperatures. The calculations results at energy
$E=E_\text{a}=1.93$~eV for magnetic fields of $B=0$, 10 and 15~T are
shown in Fig.~\ref{Temperature}. The effect of the temperature is
due to the increase of the bright exciton state population with
increasing temperature.

Constructing the time-dependent intensities $I(E,t)$ obtained from
Eq.~(\ref{inten}) using the solutions of the rate equations in the
time-resolved regime for all energies $E$, it is possible to model
the time evolution of the spectrum as a whole and to describe the
time evolution of the PL maximum $E(t)$. Using the approximate
analytical solutions given by Eqs.~(\ref{DT},\ref{AT}) and
neglecting the initial population of the bright excitons in the
donor NCs we obtain the following expression:
\begin{eqnarray}
E_0-E(t)=\Delta[E(t)][1-\exp(-t\Gamma_{\rm ET}^{\rm F})] , \label{etteor}
\end{eqnarray}
where $\Delta [E(t)]$ given by Eq.~(\ref{DENT}) depends on  energy
and thus on time. In Eq.~(\ref{EQ:shift1}) used for fitting the
experimentally observed temporal shift $E(t)$ in
Fig.~\ref{fig:EvsT}, two energy shifts $\Delta E_{1,2}$ and two
characteristic shift rates $\Gamma_{\Delta E_{1,2}}$ were used. It
is clear from Eq.~(\ref{etteor}), that the time evolution of the PL
maximum can not be described by a single exponential function even
for the case when only one type of the energy transfer takes place. We
can associate the slow component $\Gamma_{\Delta E_2}$ used in the
fit in Fig.~\ref{fig:EvsT} according to Eq.~(\ref{EQ:shift1}) with
the energy transfer rate from the dark exciton $\Gamma_{\rm ET}^{\rm
F}$. The fast component $\Gamma_{\Delta E_1}$ is most probably
caused by the energy dependence of the relaxation rate from the
bright to the dark exciton states $\gamma_0$ or by the energy
transfer from the bright exciton state that is neglected in our
simulations.

\section{Discussion and Conclusions}
\label{SecVI}

From the time-resolved PL data it is clearly seen that the typical
time scales during which the energy shift of the PL maximum and the
growth of the PL intensity at the low energy part of the spectrum
take place are significantly longer than the lifetime of the bright
excitons. This fact points on the important role of the dark
exciton, but does not provide any insight on the underlying mechanisms.
According to the experimental data and modeling results the reason
of the prominent role of the dark exciton in the ET process is that
the initial optical pumping is accompanied by the relaxation process
from the bright to the dark state and by the fast energy transfer
process from the bright state. As a result, when we start
observation at the certain conditional moment $t=0$, the populations
of the dark and bright excitons are already redistributed and the
bright exciton is nearly depopulated in the donor NCs at low
temperatures.

\begin{figure}[h!]
\centering
\includegraphics[width=8 cm]{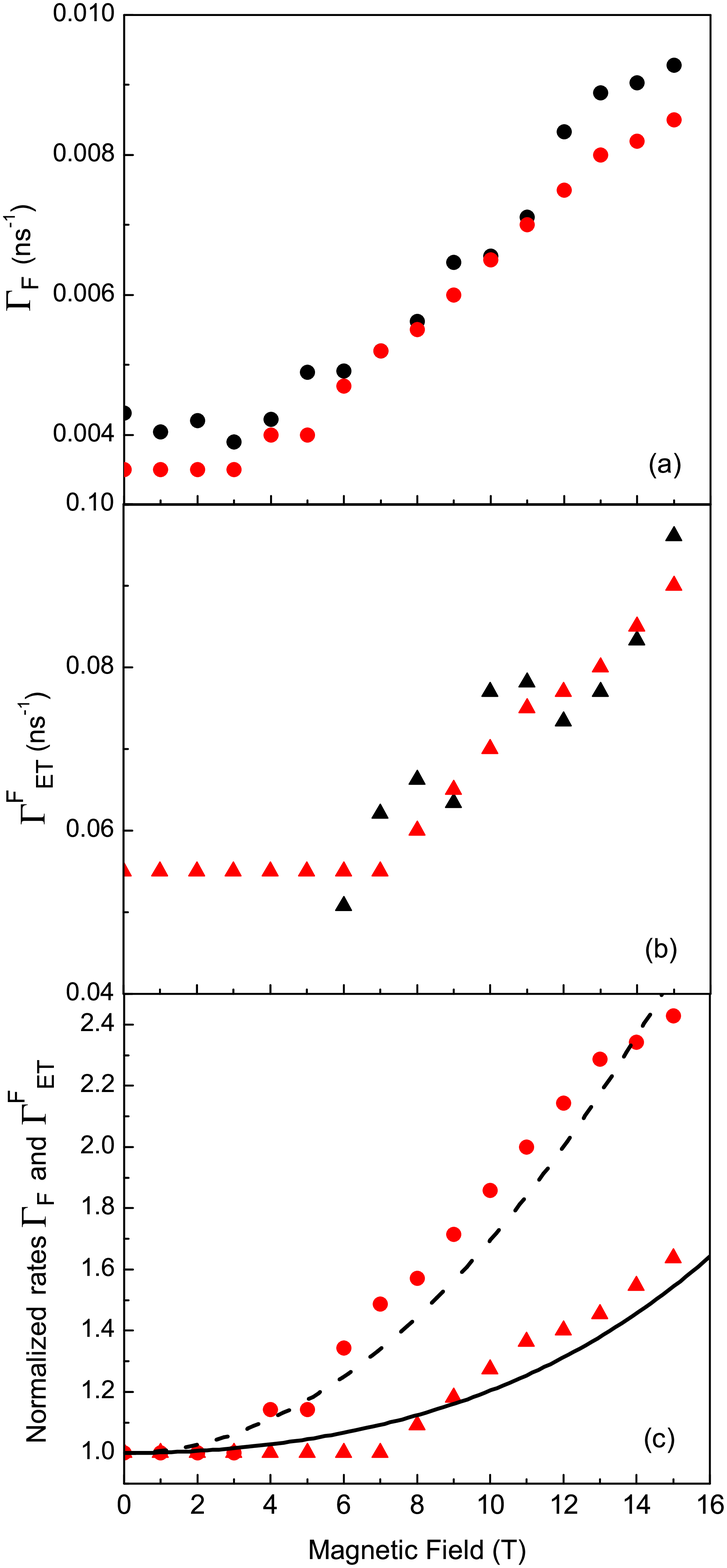}
\caption{Magnetic field dependence of (a) the radiative
recombination rate $\Gamma_\text{F}$ and (b) the energy transfer
rate of the dark exciton state $\Gamma_\text{ET}^\text{F}$,
determined from fitting the experimental decay curves at the energy
$E=E_{\rm a}=1.93$~eV using a multi-exponentional form (the black
symbols) and using the solution of the rate equations (the red
symbols). (c) Magnetic field dependence of the  normalized rates
$\Gamma_\text{ET}^\text{F}(B)/\Gamma_\text{ET}^\text{F}(0)$ and
$\Gamma_\text{F}(B)/\Gamma_\text{F}(0)$,  determined by using the
solution  of the rate equations. The dashed and solid lines give
fits with the magnetic field strength dependencies
$\Gamma_\text{F}(B)/\Gamma_\text{F}(0)=1+(B/12)^2$ and
$\Gamma_\text{ET}^\text{F}(B)/\Gamma_\text{ET}^\text{F}(0)=1+(B/12)^2+(B/24)^4$,
respectively.} \label{Rates}
\end{figure}

As mentioned above, the FRET occurs via dipole-dipole
interaction. Therefore, the efficiency of FRET depends on the
oscillator strength of the dipoles. In colloidal NCs the dark
exciton state is not completely dark  - it has some finite dipole
moment and participates in the radiative recombination even at zero
magnetic field due to the admixture of the bright exciton
state.\cite{Rodina2015} The same admixture allows the
dipole-dipole energy transfer from or to the dark exciton state in a
zero magnetic field. An external magnetic field  mixes additionally
the bright and the dark exciton states, thus increasing the dipole
moment $\mu$ of the dark exciton and its radiative
rate.\cite{Efros1996}  This additional admixture leads to the
enhancement of the energy transfer rate from the dark exciton state.

The enhancement of the energy transfer by the magnetic field can be seen already from
Fig.~\ref{fig:EvsT}(b). The shift of the peak position during the first 70~ns at $B=15$~T
(51~meV) is larger than that at 0~T (40~meV). Since the shift of the peak position is
related to the energy transfer, we can conclude that the energy transfer is enhanced by
the magnetic field. The importance of the energy transfer from the dark exciton state and
its acceleration in the magnetic field is directly demonstrated by the simulation of the
PL decay curves for a pair of donor-acceptor NCs. By adjusting the rates of excitons for
a better matching of the simulated decay curves with the experimental dependencies of
these rates on magnetic field, refined rates were achieved.

In Fig.~\ref{Rates} the
dependencies of the recombination rate $\Gamma_\text{F}$ and the ET rate
$\Gamma_\text{ET}^\text{F}$ on magnetic field are presented. The black circles show the
rates obtained directly from the multi-exponential fitting and the red circles give the
rates obtained from fitting the experimental PL decays with the solution of the rate
equations. One sees that the recombination rates obtained by both methods almost
coincide. From the results of the modeling a significant correction to magnetic field
dependence of the $\Gamma_\text{ET}^\text{F}$ rate (red triangles versus black triangles
in Fig.~\ref{Rates}(b)) in low magnetic fields is obtained. According to the modeling of
the PL decays of donor and acceptor NCs, the ET rate is constant in magnetic fields up to
6~T (the red triangles in Fig.~\ref{Rates}(b)). The magnetic field dependencies of the
normalized rates are fitted by $\Gamma_\text{F}(B)/\Gamma_\text{F}(0)=1+(B/12)^2$ and
$\Gamma_\text{ET}^\text{F}(B)/\Gamma_\text{ET}^\text{F}(0)=1+(B/12)^2+(B/24)^4$. The
$B^2$ dependence for the dark exciton recombination rate corresponds to the linear in $B$
increase of the dark exciton dipole moment $\mu \propto B$, caused by the
magnetic-field-induced admixture of the bright exciton, so that $\Delta
\Gamma_\text{F} \propto \mu^2 \propto B^2$. When the FRET occurs from the dark to the
bright exciton state, its rate will be enhanced by the magnetic field according to
Eq.~(\ref{EQ:ET}) as $\propto \mu_\text{d}^2 \propto B^2$. In the case of FRET between
two dark exciton states, its rate will be enhanced by the magnetic field according to
Eq.~(\ref{EQ:ET}) as $\propto \mu_\text{d}^2\mu_\text{a}^4 \propto B^4$. Therefore, the
$B^2$ and $B^4$ dependencies for the energy transfer rate show that the energy transfer
from the dark exciton state in the donor NC may take place to both bright and dark
exciton states in the acceptor NC.

The rate of the energy transfer from the dark exciton state
$\Gamma_\text{ET}^\text{F}(B)$ is enhanced in the magnetic field. This
enhancement is evidenced in the time-resolved studies of the PL
dynamics and the time evolution of the PL maximum in the ensemble
(see Fig.~\ref{fig:EvsT}(b)). However, the shift of the PL maximum
in the CW spectrum, $E_0-E_m$,  does not change in the external magnetic
field. This fact is related to the nearly constant value of the
energy transfer efficiency $K_{\rm ET}$ in the magnetic field and to the
weak nonlinear dependence of $E_0-E_m$ on $K_{\rm ET}$ in the range
of large $K_{\rm ET}$ values (see Fig.~\ref{shift}(b)). The effect
of the magnetic field on the CW spectrum might become more
significant in the range with smaller $K_{\rm ET}$ values.

It is worth to remind that the energy transfer rate $\Gamma_{\rm ET} \propto (R_0/R_{\rm
da})^{6}$ (see Eq.~(\ref{EQ:ET})) decreases as the sixth power of the distance between
the donor and acceptor NCs. Here $R_0$ is the characteristic F\"{o}rster radius,
corresponding to the FRET efficiency $K_{\rm ET}=(R_0/R_{\rm da})^{6}/[1+(R_0/R_{\rm da})^{6}]=0.5$. That is why the energy transfer
efficiency  is large in the
areas with high density of NCs and decreases strongly with decreasing NC density.
Therefore the areas with higher integral PL intensity correspond to areas with larger $K_{\rm ET}$ values and vice versa. The calculated dependence of the PL maximum shift
$E_m-E_0$ on $K_{\rm ET}$ in Fig.~\ref{shift}(b) thus explains the correlation between
the value of the PL maximum shift and the integral PL  intensity. For
example, from the  PL maximum shift observed in two areas (see Fig.~\ref{fig:compare}) we
can estimate the change of $K_{\rm ET}$ from 0.9 in the high density area to 0.5 in
the low density area. This corresponds to the ratio of the  $R_{\rm da}$ values in the two
areas of about 1.5. In the case when the NCs form only one layer on the substrate, this would correspond to the PL intensity ratio between the high and low density areas of about 2.25. This agrees well with the 2.5 times ratio shown in Fig.~\ref{fig:compare}(a). Exploring further the modeling assumption $R_{\text {da}} \approx d$ in the high density area, we can estimate the F\"{o}rster radius in the ensemble of the CdTe NCs as $R_0 \approx 5 - 6$~nm.

The developed theoretical model allows us to separate easily the
energy region from which the initial excitation is effectively
transferred and the energy region to which this excitation is
transferred, by analyzing the properties of the function $T_{\rm
d}(E)$ constructed from the initial distribution function $N^0(E)$
and the donor probability function $f_{\rm d}(E)$ according to
Eq.~(\ref{eq27}). For the parameters used in Fig.~\ref{shift},
$T_{\rm d}(E)<0$ for $E < 2.02$~eV. Therefore, the effect of the
additional PL rise caused by the energy transfer might be observed
already for energies below 2.02~eV. Indeed, application of the
external magnetic field allows us to observe this effect not only at
lower energy part of the spectrum, but also at the centrum of the CW
spectrum at 1.99~eV as can be seen in Fig.~\ref{fig:first50ns}(e).

In conclusion, CdTe colloidal NCs have been studied by time-resolved
photoluminescence in external magnetic fields. We prove that the
spectral diffusion observed in emission spectra is induced by the
F\"orster energy transfer. The energy transfer rate
$\Gamma_\text{ET}$ of an ensemble of randomly oriented CdTe NCs can
be enhanced by a magnetic field. The fast relaxation of excitons
from the bright to dark state as well as the admixing of the bright
to the dark exciton states caused by the magnetic field results in a
dominant role of the dark excitons in the FRET at low temperatures.

\acknowledgments The authors are thankful to R. A. Suris, Al. L.
Efros and A. N. Poddubny  for helpful discussions and to D. N.
Vakhtin for the help with developing the $C^{++}$ code. The work was
partly supported by the Deutsche Forschungsgemeinschaft and the
Russian Foundation of Basic Research in the frame of the ICRC TRR
160, by the Mercator Research Center Ruhr, by the Russian Foundation
for Basic Research (Grant No. 13-02-00888), by the Government of
Russia (project number 14.Z50.31.0021, leading scientist M. Bayer), and by the
Research Grant Council of Hong Kong S.A.R. (project T23-713/11).

%\bibliography{ET,Nanocrystals,hotrelaxation}

 \section*{\label{ApA} Appendix: Fitting function for the PL decay}

If the optical excitation with a $\delta(t)$-shaped short pulse
creates at $t=0$ the populations $N_i$ ($i=1,...,n$) in the $n$
exciton states, the resulting PL decay  with time is given by the
multi-exponential function $I_\delta(t)=\sum_i I_i(t)=\sum_i
\frac{C_i}{\tau_i} \exp(-\frac{t}{\tau_i})$, where $\tau_i$ is the
characteristic decay time, and the amplitudes $C_i$ are proportional
to the quantum efficiency of the $i$-th exciton state.

The shape of the laser pulse exciting the NCs in our experiment was
fitted by a Gaussian function
\begin{eqnarray}
I_\text{pulse}(t) = \frac{1}{\sigma_0 \sqrt{2\pi}} \exp \left[-\frac{(t - t_0)^2}{2\sigma_0^2}  \right] \, ,
\label{EQ:puls}
\end{eqnarray}
with $\sigma_0=0.34$~ns.

We assume at least $n=6$ exciton states contributing to the observed
PL (among them one is the upper excited state pumped by the laser
pulse and one is the long-living trap state giving the background
tail). The fitting function for the  PL decay was obtained as the
result of the following convolution:
\begin{eqnarray}
I_\text{PL}(t)=\int_{-\infty}^{t} I_\text{pulse}(T)I_\delta(t) dT = \nonumber \\
  \sum_{i=0}^{5}\frac{C_i}{\tau_i}\int_{-\infty}^{t} I_\text{pulse}(T) \exp (- (t-T)/\tau_i )dT  = \nonumber \\
\sum_{i=0}^{5} \frac{A_i}{2\tau_i} \exp (-\frac{t}{\tau_i}) \left[1+\text{Erf} \left( \frac{t-t_0}{\sqrt{2} \sigma_0} - \frac{\sigma_0}{\sqrt{2}\tau_i}\right)    \right] \, ,
 \label{EQ:fit}
\end{eqnarray}
where $\text{Erf}(x)$ is the error function and $A_i=C_i
\exp(1+{\sigma_0^2}/{2\tau_i^2})$.

In the fitting, the negative amplitude $A_1$ corresponding to the
initial fast rise of the PL intensity, and $1/\tau_1=13$ ns$^{-1}$
were fixed to the same values for all decay curves. The small
$1/\tau_6 <0.005$ ns$^{-1}$ was fixed for each decay curve
individually to fit the tale of the curve in the range
$I_\text{PL}(t)/I_\text{PL}(0) < 0.001$. The time $t=0$ was set for
each curve to correspond to the PL maximum, $t_0 <0$ ($|t_0|<1$ ns)
and $I_\text{PL}(t_0)=0$. The four rates $1/\tau_i>0$ and the
amplitudes $A_i$ ($i=2,3,4,5$) varied within the fitting
procedure which was performed with a specially developed
$C^{++}$ code. Positive amplitudes $A_i>0$ always correspond to the
decaying components, while negative amplitude $A_i<0$ allowed us to
describe the rising component in magnetic field at the low energy
side of the spectrum.

\end{document}